\documentclass[10pt,journal,letterpaper,compsoc]{IEEEtran}

\usepackage[nocompress]{cite}

\usepackage[final]{trackchanges}

\usepackage{stmaryrd}
\usepackage{wasysym}
\usepackage{latexsym}
\usepackage{amssymb}
\usepackage{eepic}
\usepackage{epic}
\usepackage{rotating}
\usepackage{bbm}

\newcommand{\proofoflemma}[2] {{{\sc Proof of  lemma}~\ref{#1}}:\/ #2 ~\hspace*{\fill}
  $\rule{1ex}{1.5ex}$\par\vspace{1pc}}

\newcommand{\proofofprop}[2] {{{\sc Proof of  proposition}~\ref{#1}}:\/ #2 ~\hspace*{\fill}
  $\rule{1ex}{1.5ex}$\par\vspace{1pc}}

\newcommand{\proofoftheorem}[2] {{{\sc Proof of theorem}~\ref{#1}
        (p.~\pageref{#1})}:\/ #2 ~\hspace*{\fill}
        $\rule{1ex}{1.5ex}$\par\vspace{1pc}}

\newtheorem{theorem}{Therem}[section]
\newtheorem{proposition}{Proposition}[section]
\newtheorem{lemma}{Lemma}[section]
\newtheorem{example}{Example}[section]
\newtheorem{definition}{Definition}[section]

\newcommand{\comment}[1]{}

\def\constraints{\mathbbm{Cnd}}

\def\events{\mathbbm{Evt}}
\def\traces{\mathbbm{Tr}}
\def\tokens{\mathbbm{Tk}}
\def\diagrams{\mathbbm{Sd}}
\def\taggedtraces{\mathbbm{Tt}}
\def\labels{\mathbbm{Lab}}
\def\names{\mathbbm{Name}}

\def\lifelines{\mathit{lifelines}}

\def\tag{\mathit{tag}}
\def\untag{\mathit{untag}}
\def\dom{\mathit{dom}}

\def\conformsto{{\rhd}}

\def\strict{{\twoheadrightarrow}}

\newcommand\sem[1]{{[\!\![#1]\!\!]}}

\newcommand\region[1]{{\llparenthesis #1 \rrparenthesis}}

\newcommand{\todo}[1]{\marginpar{\sf #1}}

\begin{document}

\bibliographystyle{plain}

\title{Required Behavior of  Sequence Diagrams: Semantics and Conformance}
\author{
 \IEEEauthorblockN{Lunjin Lu and Dae-Kyoo Kim}

\IEEEauthorblockA{
  Oakland University
} }

\maketitle


\begin{abstract}
Sequence diagrams are a widely used design notation for describing
software behaviors. Many reusable software artifacts such as design
patterns and design aspects make use of sequence diagrams to describe
interaction behaviors. When a pattern or an aspect is reused in an
application, it is important to ensure that the sequence diagrams for
the application conform to the corresponding sequence diagrams for the
pattern or aspect. Reasoning about conformance relationship between
sequence diagrams has not been addressed adequately in literature. In
this paper, we focus on required behavior specified by a UML sequence
diagram. A novel trace semantics is given that captures precisely
required behavior specified by a sequence diagram and a conformance
relation between sequence diagrams is formalized based on the
semantics. \add[LL]{Properties of the trace semantics and the
  conformance relation are studied.}
\end{abstract}


\section{Introduction}

The Unified Modelling Language (UML) sequence diagrams~\cite{UML05}
and their predecessors Message sequence charts~\cite{MSC} are
specification languages that have been widely used for specifying
interaction behaviors in  software development.  A sequence diagram (SD)
describes inter-object/inter-process behavior of a system in graphical
manner.  It shows as parallel vertical lines different objects or
processes that communicate with each other via messages that are shown
as horizontal arrows. Each message has an associated sending event and
an associated receiving event. Events are basic behavioral constructs
of UML SDs.  They can be combined to form larger behavioral constructs
called fragments. A fragment is either an event or formed of an
interaction operator, one or two operands which may be themselves
fragments and an optional condition. It involves a collection of
lifelines and is formed of events and smaller fragments. In this
paper, we shall use the terms SD and fragment interchangeably.

\begin{figure*}
\begin{center}
\scalebox{0.7}{\includegraphics{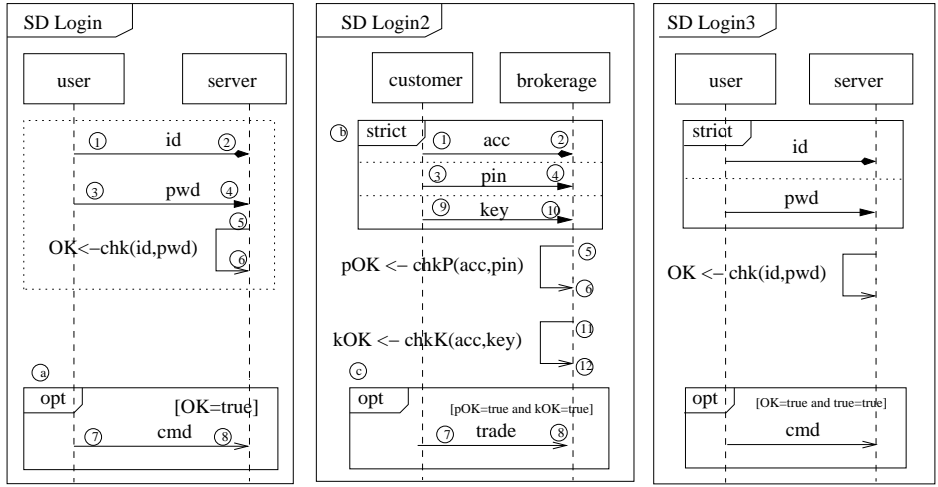}}
\end{center}
\caption{Sequence Diagrams for the Running Example}
\label{fig:login}
\end{figure*}

\begin{example}\hspace{1pc}
 \label{ex:running}
We shall use SDs in Fig.~\ref{fig:login} as a running example.  In SD
Login, the \textit{alt} fragment is labelled $a$ and the sending and
receiving events for a message are labelled with two consecutive
numbers.  Let $e_i$ abbreviate the event labelled $i$. For instance,
$e_1$ abbreviates $!id$ the sending event of message $id$ and $e_2$
abbreviates $?id$ the receiving event of message $id$ omitting the
sender and the receiver of the message.  The SD Login may be thought
of as a pattern for a user to sign in to get a service from a
server. The user provides to the server his user-id \textit{id} and
password \textit{pwd}.  The server checks if the user-id and password
are correct using a system variable \textit{OK} to indicate the
result. If \textit{OK} equals \textit{true} then the user issues a
command \textit{cmd} to the server.
\end{example}

\subsection{Motivation}

Software development can greatly benefit from reusing existing
artifacts including architectural patterns, design patterns, design
aspects, software components and code.  An important issue that arises
in reusing an artifact is how to ensure that the desirable properties
of the artifact are preserved.  This issue becomes harder and more
critical when the artifact involves significant interaction
behaviors. Many reusable artifacts make use of SDs to specify
interaction behaviors. If an artifact is reused in an application, it
is important to verify that the SDs in the application conforms to the
SDs in the artifact. Otherwise, the intended benefits of the artifact
cannot be guaranteed. A special case of reuse is refinement in which
an SD developed in an earlier stage is refined to obtain an SD in a
later stage. Software design is an iterative process. Starting with an
initial design model, a series of design models are obtained, each of
which refines its predecessor. This process is applied to behavioral
models as well as structural models.  Each immediate model needs be
verified against its predecessor.  A fundamental issue arising from
using SDs is whether one SD model correctly refines its predecessor in
that it possesses all required behaviors that are mandated by the
predecessor and at the same time rejects all proscribed behaviors that
are prohibited by its predecessor.

\begin{example}\hspace{1pc} \label{ex:behavior} Consider SD Login again. Let
$t= e_1e_2e_3e_4e_5e_6$ and $t' = e_1e_3e_2e_4e_5e_6$.  Let $r_1 =
  t[OK=true]e_7e_8$, $r_2 = t[OK=false]$, $r'_1 =t'[OK=true]e_7e_8$,
  $r'_2 = t'[OK=false]$, SD Login specifies two alternative minimum obligations
  $\mathcal{O}= \{r_1,r_2\}$ and $\mathcal{O}'= \{r'_1,r'_2\}$. A
  system satisfies SD Login if it fulfils one of the two obligations. A
  system fulfills $\mathcal{O}$ if it has runs that produce the trace
  $r_1$ and runs that produce the trace $r_2$. A system that fulfils
  $\mathcal{O}'$ can be described similarly. That guard conditions
  occur in traces shall be explained later.
\end{example}

In aspect-oriented software development, design models may be
developed by composing aspects with primary models, which involves
composing sequence and class diagrams from
aspects~\cite{FranceRGG04}. It is necessary to verify that the
composed SD conforms to each of the component SDs.  In pattern based
development, the designer needs to check if an SD developed by the
designer conforms to the behavior of a design pattern~\cite{GHJV95} in
the sense that it is a valid realization of the pattern.  The purpose
of using design patterns is to improve the quality of software
designs. However, \change[LL]{an invalid realization could deteriorate
  rather than improve its quality.}{an invalid realization could break
  the design rather than improve its quality.} Various efforts have
been made to facilitate pattern realization.  A common approach is
using templates where pattern participants are parameterized (e.g.,
see \cite{GSJ00,MHG02}). A pattern is instantiated by stamping out the
template with parameters bound to application elements.  In many
cases, instantiated pattern realizations often require significant
modifications such as adding new elements or modifying instantiated
elements to accommodate application-specific needs. Since these
activities may break pattern conformance and compromise the benefits
of using design patterns, it is imperative to check if the application
conforms to the pattern.

An SD is partial in that it describes a number of alternative
obligations that an implementation may choose to fulfil. For instance,
the fragment operator $\mathit{par}$ does not mandate that an
implementation must be distributed, concurrent or multi-threaded. It
rather indicates \add[LL]{that} the implementation can realize any interleaving of
the behaviors of its operands. When the SD is  reused, it is
made more defined in that the number of alternatives is reduced. An SD
under reuse may be undergone a numbers of changes including the
following. Firstly the names of lifelines and messages may be
changed. Such changes are necessary to avoid names conflicts or to
better reflect the developer's intention. For instance, the lifeline
\textit{user} in the SD Login may be renamed to \textit{customer} for
a business application.  Secondly, control structure of the SD may be
changed to eliminate non-determinism.  Finally,
new messages (and hence new events) may be added.

\begin{example}\hspace{1pc}
 The SD Login2 describes a sign-in interaction for a customer of a
 brokerage and can be obtained by refining SD \textit{Login} as
 follows. Firstly, the developer renames \textit{user} to
 \textit{customer}, \textit{server} to \textit{brokerage}, \textit{id}
 to \textit{acc}, \textit{pwd} to \textit{pin}, \textit{chk} to
 \textit{chkP}, \textit{OK} to \textit{pOK} and  \textit{cmd} to
 \textit{trade}.  The developer then eliminates non-determinism by
 requiring that \textit{?acc} occurs before \textit{!pin}.  He also
 introduces a new system variable \textit{kOK} and two new messages
 \textit{key} and \textit{chkK} which produces output
 \textit{kOK}.  The condition for the \textit{opt} fragment
 has also been strengthened.

That SD Login2 conforms to SD Login can be informally checked as
follows.  SD Login3 may be obtained from SD Login2 by hiding messages
\textit{key} and \textit{chkK}, using default value \textit{true} for
\textit{kOK} and changing the names back. Moreover, SD Login3 is same as
SD Login except that in SD Login3, ?id must occur before !pwd while they can
occur in any order in SD Login.  Formally, SD Login3 specifies one
obligation which is $\mathcal{O}$ given in Example~\ref{ex:behavior}.
Any system satisfying SD Login3 fulfils $\mathcal{O}$ - one of the two
alternative obligations of SD Login.  SD Login2 conforms to SD Login because
SD Login3 is obtained from SD Login2 by renaming and hiding and it specifies
$\mathcal{O}$.
\end{example}

The above example illustrates conformance checking. In conformance
checking, an SD is verified to conform to another with respect to a
set of unobservable events $\mathcal{U}$ and a mapping $\rho$ that
changes the names of system variables, lifelines and messages and
assigns default values to some system variables. Conformance inference
on the other hand infers automatically possible $\mathcal{U}$ and
$\rho$ with respect to which an SD conforms to another. Conformance
inference requires a formalization of a conformance relation between
SDs which in turn requires a formal trace semantics that captures
precisely behavior of SDs.

\subsection{Contributions}

In the existing trace
semantics~\cite{Storrle03,CengarleK04,HaugenHRS05}, an SD denotes
a set of all possible traces that the specified system may produce
and a set of proscribed traces that the specified system must not
produce. They are useful as a semantic base for verifying SDs
against safety properties. However, they are not useful as a
semantic base for a conformance relation between SDs since they do
not tell which possible traces are required in that the specified
system must produce.  As a special case of conformance, refinement
has been studied for statecharts and modal transition systems.
However, translations from SDs to these state machine models
either have not been proved correct with respect to a formal trace
semantics or introduce behaviors that are not required by SDs.
Thus, results on refinement of these state machine models do not
carry over to SDs. Related work will be discussed in
Section~\ref{sec:related}.

In this paper, we give a trace semantics that characterizes required
behavior specified in an SD and formalize a conformance relationship
between SDs.  Conformance is defined in terms of a simulation relation
between traces. The notion of one trace simulates another will be made
clear later.  Roughly speaking, a trace $t_1$ simulates another trace
$t_2$ if all events in $t_2$ are simulated in $t_1$ in order in which
they occur and there are no observable events in $t_1$ other than
those that simulate events in $t_2$. An SD $D_1$ refines another $D_2$
if an implementation of $D_1$ is also an implementation of $D_2$. In
other words, $D_1$ preserves required behavior of $D_2$ but may
specify more required behaviors. An SD $D_0$ conforms to another $D_2$
if there is an SD $D_1$ such that $D_1$ refines $D_2$ and $D_1$ can be
obtained from $D_0$ by renaming lifelines and messages, hiding events
and assigning values to system variables. These concepts will be made
clearer in Section~\ref{sec:preservation}.

The main contributions of this work are as follows.
\begin{itemize}

\item A novel trace semantics is formulated for a subset of UML SDs.
  \add{Unlike the trace semantics proposed in literature
    \cite{CengarleK04,HaugenHRS05,Storrle03} that capture possible
    behavior of SDs, our trace semantics captures precisely required
    behavior of SDs and forms a basis for a semantics based conformance
    relation. As discussed in Section~\ref{sec:related}, a conformance
    relation should not be based on a semantics for possible behavior
    of SDs.} \add{ While those trace semantics for possible behavior
    of SDs ignore guard conditions, our trace semantics encodes guard
    conditions in SDs as elements of traces. This is required to
    ensure soundness of conformance, as discussed in
    Section~\ref{sec:related}.}  The semantics possesses
  substitutivity which is not enjoyed by trace semantics proposed in
  literature~\cite{CengarleK04,HaugenHRS05,Storrle03}. A nice
  consequence of substitutivity is that a component of an SD can be
  replaced with a semantically equivalent component without changing
  the semantics of the SD.

\item A conformance relation between SDs is defined based on the
  semantics. A desirable property of the conformance relation is that
  it allows messages and lifelines to be renamed during conformance.
  \add{The conformance relation is transitive, implying that
    conformance can be verified in step wise manner.}

\end{itemize}

The rest of the paper is organized as follows.
Section~\ref{sec:related} discusses about related work.
Section~\ref{sec:syntax} presents an abstract syntax for SDs and
Section~\ref{sec:semantics} defines the trace semantics.
Section~\ref{sec:preservation} defines the conformance relation and
Sections~\ref{sec:case} and \ref{sec:case2} present two case
examples. Section~\ref{sec:conclusion} concludes. Proofs are placed in
an appendix. This paper is an extension of \cite{LuK:ICECCS11}. The
conformance relation presented in this paper generalizes the
refinement relation in \cite{LuK:ICECCS11} by taking into account
event hiding, renaming of lifelines and messages and assignment of
values to system variables.  The extension includes the conformance
relation in Section 5.3, two case examples in Sections 6 and 7, and
proofs in the appendix.  Section 2 is written to include more detailed
discussion of related work. Other sections are written to include more
examples and to improve presentation.

\section{Related Work} \label{sec:related}
We shall now put our work in the context of the existing work.  Since
our work is concerned with semantic-based conformance reasoning
for SDs, we focus on observational semantics of
and conformance/refinement relations on SDs and their variants.
For a survey on semantics of SDs, see \cite{MicskeiW2010}.

\subsection{Syntactic-based Refinement and Conformance}
Mauw and Reniers propose instance refinement for
Interworkings~\cite{MauwR96}. Interworkings are similar to MSCs except
that messages in Interworkings are synchronous and have only two
interaction operators: \textit{seq} and \textit{par}. When an instance
is refined, it is decomposed into several component instances and new
messages may be added between these component instances. \comment{The
  semantics is defined by translating an interworking into a process
  algebra term and the refinement is defined in terms of bisimulation
  on process algebra terms.}  Engels~\cite{Engels98} studies message
refinement for basic MSCs (bMSCs) which are MSCs without interaction
operators. A message $m$ in a bMSC $k$ is refined by another bMSC $p$
which has two distinct instances $s$ and $r$ corresponding to the
sender and the receiver of $m$ respectively. The refined bMSC, denoted
$k[p/m]$, is obtained by removing $m$ and splicing $p$ into $k$ such
that orders on events imposed by $k$ and $p$ are preserved. In
addition, any event in $k$ preceding $!m$ now precedes all sending
events on $s$ and any receiving event on $r$ now precedes all those
events that follow $?m$ in $k$.  Muscholl et
al. \cite{MuschollPS98,MuschollP00} call a bMSC M to match another N
with respect to a set of messages T if M can be obtained from N by
removing zero or more messages in T. The matching relation is extended
to hierarchical MSCs (HMSCs) which are automata with bMSCs as
transitions.  An HMSC H matches another K if there is a pair $\langle
p_1, p_2\rangle$ of paths of H and K such that $b_1$ matches $b_2$
where $b_i$ is sequential composition of all bMSCs along $p_i$ for
$i=1,2$.  Khendek et. al \cite{Khendek:2001:SDM} propose a notion of
conformance for MSCs.  A bMSC $M_2$ conforms to another bMSC $M_1$ if
$M_2$ can be obtained from $M_1$ by refining one or more instances in
$M_1$ and adding new messages between new and/or existing
instances. The conformance relation is extended to HMSCs that are
sequential compositions of bMSCs. The HMSC $seq_{i\in I} M_i$ conforms
to $seq_{j\in J} N_j$ if there is an $M_i$ conforming to $N_j$ for
each $j\in J$. The above notions of refinement, matching and
conformance are syntactic-based in that they are decomposing,
introducing and removing instances and messages. They are defined for
a subset of both MSCs and SDs.  While they represent some reuses of
MSCs, they are restrictive. For instance, they do not always allow us
to replace one MSC with a semantically equivalent MSC. Thus, a notion
of conformance based on semantics is needed to allow more flexible
reuse.

\subsection{Direct Style Semantics and Refinement}
There is little work on semantic-based conformance in
general. However, semantic-based refinement which is a special case of
semantic-based conformance has attracted much attention.  In
\cite{CengarleK04,CengarleGW06} and \cite{Storrle03}, the semantics of
an SD is a pair consisting of a set of positive traces and a set of
negative traces.  Haugen et\ al.~\cite{HaugenHRS05} define the
semantics of an SD as a set of obligations all of which must be
fulfilled.  Each obligation is a pair consisting of a set of positive
traces and a set of negative traces. Without the fragment operator
{\it xalt} which they introduce to capture the required
non-determinism, the semantics of an SD contains a single obligation
and is equivalent to that of \cite{Storrle03}.  Lund and St{\o}len
provide an operational semantics for UML SDs~\cite{LundS06} which is
sound and complete with respect to the trace semantics of
\cite{HaugenHRS05}. There is no discussion on refinement in
\cite{LundS06}.  Refinement in
~\cite{CengarleK04,HaugenHRS05,Storrle03} is defined as eliminating
positive traces and making them proscribed. Under this interpretation,
the system is only required to have one of positive traces, which is
problematic as shown below. For SD Login, the set of positive traces
is $\{te_7e_8,t'e_7e_8,t,t'\}$ where $t$ and $t'$ are given in
Example~\ref{ex:behavior}. \comment{Whilst the set of negative traces
  captures precisely bad behavior proscribed by SD Login, } The set of
positive traces does not capture precisely required behavior of SD
Login.  As shown in Example~\ref{ex:behavior}, the specified system
does not need to produce all positive traces in order to satisfy SD
Login. It only has to produce $t$ and $te_7e_8$ or $t'$ and
$t'e_7e_8$.

 A logical semantics for basic SDs is presented in \cite{ChoKCB02}.  A
 basic SD $D$ has only finite number of finite traces.  The semantics
 of $D$ is a temporal logic formulae with freeze
 quantifier~\cite{ChoKCB01}. The semantics captures a single set of
 possible traces and applies to a small subset of SDs.  SDs are
 formalized in ~\cite{Aredo02} as PVS theories that specify a set of
 possible traces for each object in the system. Refinement is not
 discussed in \cite{Aredo02,ChoKCB02}.

The above trace
semantics~\cite{Aredo02,CengarleK04,ChoKCB02,HaugenHRS05,Storrle03}
associate an SD with a set of possible traces. They are useful for
verification of SDs against safety properties of SDs such as dead-lock
freedom~\cite{AlurY99}. \comment{Their approach is to translate an SD
  $D$ to a state machine $A$ that accepts all possible traces and
  express a safety property as another state machine $B$ that accepts
  all bad traces. The SD is verified if $L(A)\cap L(B)=\emptyset$
  where $L(M)$ is the language accepted by the state machine $M$.}
However, they are inadequate for SD conformance reasoning in several
aspects.  Firstly, they do not distinguish required behaviors from
optional behaviors as pointed out in \cite{SenguptaC06}.  Secondly,
they ignore guard conditions, which compromises soundness of
conformance reasoning. For instance, let $D_1=alt(c,!m,!n)$ and
$D_2=!m$ then ignoring constraints would assign two traces !m and !n
to $D_1$ and one trace !m to $D_2$ and lead to a false conclusion that
$D_1$ possesses all required behaviors of $D_2$. In fact, $D_2$
requires the specified system to produce $!m$ in all runs whilst $D_1$
only requires the specified system to produce $!m$ in those runs that
starts with system states in which the condition $c$ holds.  Thirdly,
they do not deal with critical regions adequately. All but one
\cite{Storrle03} of above mentioned semantics are defined for SDs with
critical regions.  Semantics in \cite{Storrle03} does not possess
substitutivity in the presence of critical regions. Let $D_1$ be
critical(strict(!a,!b)) and $D_2$ strict(!a,!b). Then $D_1$ and $D_2$
have the same meaning according to \cite{Storrle03} but par($D_1$,!c)
and par($D_2$,!c) do not.

The semantics in \cite{LiLJ04} captures the effect of a synchronous
message specified by an SD on logical properties of the specified
system. It abstracts away too much details of interactions and hence
is not amiable to analysis of trace properties including behavioral
conformance.  The same applies to logical semantics for MSCs in
\cite{Broy05}.

\subsection{Translation to Automata and Process Calculi}
SDs and their predecessor MSCs have been studied via translation to
automata, process calculi and other formalisms.  Mauw and Reniers
\cite{MauwR94} present a process-based semantics for basic MSCs (in
short bMSCs) which are MSCs without fragment operators.  A bMSC is
translated to a process in ACP~\cite{BergstraK85}. \comment{They show
  that any trace of a bMSC is a sufficient representation of the bMSC
  because all instances in the bMSC can be reconstructed from the
  trace.} Chen et\ al. provide semantics for bMSCs with
data~\cite{ChenKS05} by translating a bMSC to a process in a variant
of CCS~\cite{Milner}.

Whittle and Schumann generate statecharts from a collection of UML SDs
and a collection of OCL constraints~\cite{WhittleS00}.  Ziadi
et\ al. translate a scenario specification in UML SDs into
statecharts~\cite{ZiadiHJ04}. As noted in~\cite{ZiadiHJ04}, such
translations result in statecharts whose behaviors include all
behaviors of the scenario but include behaviors that are not required
by the scenarios.  Hammal defines the semantics of an SD as an
automaton whose states are maps from objects to traces and whose edges
are labelled with events~\cite{Hammal06}.  To obtain a finite
automaton, possible traces that contain the same set of events are
identified.  An SD has also been translated to a Petri net (e.g.,
\cite{CardosoS01,EichnerFMSS05,Fernandes07}) with lifelines translated
to processes, actions to transitions and messages to communication
places and to abstract state machines (e.g.,
\cite{CavarraK04,KohlmeyerG09,Xiang:2009}).  Refinement is not
considered in
\cite{CardosoS01,CavarraK04,EichnerFMSS05,Fernandes07,Hammal06,KohlmeyerG09,WhittleS00,ZiadiHJ04,Xiang:2009}.

Grosu and Smolka give safety and liveness semantics for SDs in terms
of B\"{u}chi automata~\cite{GrosuS05}. Refinement is defined  as
set containment.  Knapp et al. \cite{KnappW06} translate SDs to
automata for model checking using the SPIN model checker. Alur
et\ al. translate MSCs to automata for checking against safety
properties such as dead-lock
freedom~\cite{AlurY99,AlurEY03}. \comment{Their approach is to
  translate an SD $D$ to a state machine $A$ that accepts all possible
  traces and express a safety property as another state machine $B$
  that accepts all bad traces. The SD is verified if $L(A)\cap
  L(B)=\emptyset$ where $L(M)$ is the language accepted by the state
  machine $M$.} Refinement is not discussed in
\cite{AlurEY03,AlurY99,KnappW06}.

 Uchitel et\ al. \cite{UchitelKM03}
synthesize a labelled transition system (LTS) from MSC scenarios and
use it to detect scenarios that are implied by positive and negative
scenarios \cite{UchitelKM04}. An LTS is a finite state machine with
each transition labelled with an action (event) or $\tau$.  In
\cite{UchitelBC09}, modal transition systems are synthesized from
properties in Fluent Linear Temporal Logic~\cite{GiannakopoulouM03}
and traces of scenarios.  A modal transition system
(MTS)~\cite{LarsenT88} is a generalization of an LTS.  An MTS has two
transition relations, one describing possible transitions and the
other required transitions.  Possible transitions that are not
required can be made required or proscribed in later phase of model
development.  Sibay et\ al. ~\cite{SibayUB08} translate existential
LSCs to MTSs.  Krka et\ al. synthesize MTSs from a set of basic SDs and
OCL constraints ~\cite{KrkaBEM09} - one MTS for each component of the
specified system. Refinement of MTSs has been studied in
\cite{FischbeinBU09,FischbeinU08,FischbeinUB06}.

Defining semantics of SDs via translation allows us to leverage
established results in other areas to analyze SDs.
Bisimulation~\cite{FischbeinBU09,FischbeinU08,FischbeinUB06,Park81},
\textit{must} preorder~\cite{Hennessy85,Hennessy88,NicolaH84} and
\textit{failures} preorder~\cite{Hoa85} are close
relatives~\cite{CleavelandH93,EshuisF02} and have been used to define
refinement of automata and processes.  Refinement in bisimulation,
\textit{must} and \textit{failures} preorder semantics keeps required
traces while decreasing non-determinism.  \comment {An automaton $M$
  \textit{must} accept a trace $e_1e_2\cdots e_n$ if $M$ accepts
  $e_1e_2\cdots e_n$ no matter how $M$ chooses its transition on each
  input symbol $e_i$. The \textit{must} preorder is denoted
  $M_1\sqsubseteq M_2$ and holds if $M_2$ must accepts all traces
  $M_1$ must accept and $M_1$ and $M_2$ accept the same language.
  When $M_1\sqsubseteq M_2$ holds, $M_2$ is a refinement of $M_1$ in
  the sense that it is more deterministic than $M_1$.}  However, the
translation algorithms are limited to small subsets of SDs and ignore
essential features of SDs. For instance, they all ignore critical
regions and they all except \cite{KnappW06} ignore guard
conditions. Note that guard conditions cannot be disregarded for
conformance reasoning as pointed out in the previous section. It is
difficult to extend these translation algorithms to include critical
regions.

\subsection{Other Extensions to MSCs and SDs}
There have been work on empowering SDs and MSCs with more language
constructs. We now briefly discuss those extensions that are more
influential. Live Sequence Charts (LSCs)~\cite{DammH01,Harel:2003} are introduced
to capture existential and universal modalities.  \comment { Basic LSC
  charts are similar to basic MSCs.  An existential LSC (eLSC) is a
  basic LSC surrounded by a dashed rectangle and a universal LSC
  (uLSC) is a pair of basic LSCs called pre-chart and main chart
  respectively.  A basic LSC is interpreted as a set of traces. The
  semantics of an eLSC with basic chart $B$ consists of all the traces
  $w$ such that $w=w_1w_2w_3$ and $w_2$ is in the semantics of $B$ for
  some $w_1,w_2$ and $w_3$ \cite{Bontemps:2005:LSC}.  Semantics of a
  uLSC with pre-chart $P$ and main chart $M$ contains a trace $w$ if
  and only if either $w$ does not have a sub-trace in the semantics of
  $P$ or for all $w_1,w_2,w_3$ such that $w=w_1w_2w_3$ and $w_2$ is in
  the semantics of $P$, a prefix of $w_3$ is in the semantics of
  $M$. LSCs has no formal notion of refinement, as pointed out in
  \cite{SeehusenSS09}. } LSCs have been subject to as much study as
MSCs and SDs (see references in \cite{Bontemps:2005:LSC,HarelMS08,Harel:2003})
and have had synergistic impact on SDs.
Modal Sequence Diagrams (MSD) \cite{HarelKM07,HarelM08} are an
extension of SDs to deal with challenges with \textit{negate} and
\textit{assert} operators.  In MSD, a modal stereotype is attached to
an interaction fragment to specify whether it describes a hot
(required) or cold (possible) behavior.

Triggered MSCs (TMSCs)~\cite{SenguptaC06} are introduced to catch
conditional scenarios. Each instance in a TMSC has a trigger and an
action. A system satisfies a TMSC if whenever it exhibits the behavior
described by the trigger of an instance, its subsequent behavior is
limited to the behavior described by the action of the instance. The
meaning of a TMSC is an acceptance tree~\cite{Hennessy85,Hennessy88}
which maps a trace $w$ to an acceptance set which is a measure of
non-determinism of the system after exhibiting $w$.  The \textit{must}
preorder has been adapted for Triggered MSCs
(TMSCs)~\cite{SenguptaC06}.  TMSCs do not have a construct similar to
$\textit{critical}$ or use guard conditions \cite{SenguptaC06}.

\subsection{Summary}
In summary, a notion of semantics-based conformance is needed to allow
more flexible reuse of SDs, which requires a formal semantics.  There
does not exist a suitable semantics for conformance reasoning because
direct style semantics do not precisely capture required behaviors by
SDs and translations to other formalisms disregard essential features
of SDs.

\section{Abstract Syntax} \label{sec:syntax}

An SD specifies runtime behavior of a system in a
graphical manner. It shows as parallel vertical lines different
objects or processes that communicate with each other via messages
that are shown as horizontal arrows. A simple diagram which does not
have any combined fragment has been modelled as a partial order on
event occurrences~\cite{CardosoS01}.
Intuitively, $e_1\leadsto e_2$ indicates that $e_1$ occurs no later
than $e_2$.  Since $\leadsto$ is asymmetric, there is unique
irreflexive and non-transitive relation $\strict$ such that
$\leadsto=\strict^{*}$ where $*$ is the reflexive and transitive
closure operator. The relation $\strict$ is the transitive and
reflexive reduction of $\leadsto$ and we call it a strict
sequencing order.

Events are basic behavioral constructs of UML SDs.
They can be combined to form larger behavioral constructs called
fragments. A fragment is formed of an interaction operator, one or
two operands which may be themselves fragments and an optional
condition. It involves a collection of lifelines and is formed of
events and smaller fragments.  In this sense, an event is a
primitive fragment.

In this work, we do not consider interaction operators {\em ignore,
  consider, assert}, {\em neg} or {\em break}.  Since we are concerned
with checking if the behaviors described by one model are found in
another model, {\em ignore} and {\em consider} fragments play no role
and thus can be removed.  Despite prior effort in clarifying {\em
  assert} and {\em neg} operators~\cite{HarelM08,Storrle:assert}, no
commonly accepted interpretation for these operators has been
established. The UML 2.0 standard states that a {\em break} fragment
is a breaking scenario that is performed instead of the remainder of
its enclosing fragment. It is not clear whether the enclosing fragment
means the innermost enclosing fragment or the innermost loop fragment.
We assume that all references to SDs through interaction operator {\it
  ref} have been eliminated via syntactic unfolding since SDs are
non-recursive.

Let $\names$ be a denumerable set of names of messages, lifelines,
system variables and values. \add[LL]{An event $e$ in $\events$ has the
  following structure.} An event sending a message with name
$N\in\names$, sender $S\in\names$, receiver $R\in\names$,
parameter list \change{$P\in\wp(\names)$}{$P\subseteq \names$} is
written as $(!,N(P),S,R)$ or $!N(S,R,P)$, and the corresponding
receiving event $(?,N(P),S,R)$ or $?N(S,R,P)$. We shall simply
write $!N(P)$ or $?N(P)$ when the sender and receiver are clear
from context. \remove[LL]{Application of a function
  $\rho:\names\mapsto\names$ to a syntactic object $o$, denoted
  $\rho(o)$, is obtained from substituting $\rho(n)$ for each
  occurrence of $n$ in $o$.}  We abstract from details of guard
conditions \add[LL]{$c$ in $\constraints$} and \change[LL]{requires two operations
  on guard conditions $\neg_{c}$ and $\vee_{c}$ with the following
  properties: $\neg_{c}(c)$ is a guard condition and it is true in a
  value assignment iff $c$ is false in the same value assignment; $c_1
  \vee_{c} c_2$ is true in a value assignment if either $c_1$ or $c_2$
  or both are true in the value assignment}{require that the
  collection of guard conditions are closed under classical logical
  negation ($\neg_{c}$), conjunction ($\wedge_c$) and disjunction
  ($\vee_{c}$) operations}. We write $c_1\models_{c} c_2$ iff $c_2$ is
true in all value assignments in which $c_1$ is true.  Other primitive
syntactic entities are labels $\ell$ in $\labels$ and $\tau$
representing unobservable events. The abstract syntax for SDs in
$\diagrams$ is given below.

\[ \mathit{D} :: =
\begin{array}[t]{l}
~~\tau \mid e \mid opt(c,D_1) \mid
  alt(c,D_1,D_2)  \mid loop(c,D_1) \\
 \mid  critical(D_1)  \mid
  par(D_1,D_2)  \mid strict(D_1,D_2)
\\ \mid  seq(D_1,D_2) \mid block(L,\iota, \strict)
\end{array}
\]
where the interaction operator {\it block} is introduced to structure
operands of other interaction operators, $L$ is a non-empty set of
labels, $\iota$ is a mapping from $L$ to $\diagrams$, $\strict$ is an
irreflexive and non-transitive relations on $L$ such that
$\strict^{*}$ is a partial order. \add{The mapping $\iota$ associates
  each label in $L$ with an SD}. $\langle L,\iota,\strict^\ast\rangle$
is a partially ordered multiset~\cite{POMSET}.

\begin{example}\hspace{1pc}  \label{ex:running:syntax}
The SD Login in Fig.~\ref{fig:login} is represented in the abstract
syntax as \( \mathit{Login} = block( \{ 1..6, a\}, \{i\mapsto e_i\mid
1\leq i \leq 6\} \cup\{a\mapsto D_a\}, \strict_0)\) where
$\strict_0=\{\langle 1,2\rangle, \langle 1,3\rangle, \langle
3,4\rangle,  \langle 2,4\rangle,\langle 4,5\rangle, \langle 5,6\rangle,
\langle 6,a\rangle\}$ and $D_a= opt(OK=true, block(\{7,$ $8\},
\{7\mapsto e_7, 8\mapsto e_8\}, \{\langle 7,8\rangle\}))$.
\end{example}

\begin{figure*}
\begin{center}
\scalebox{0.45}{\includegraphics{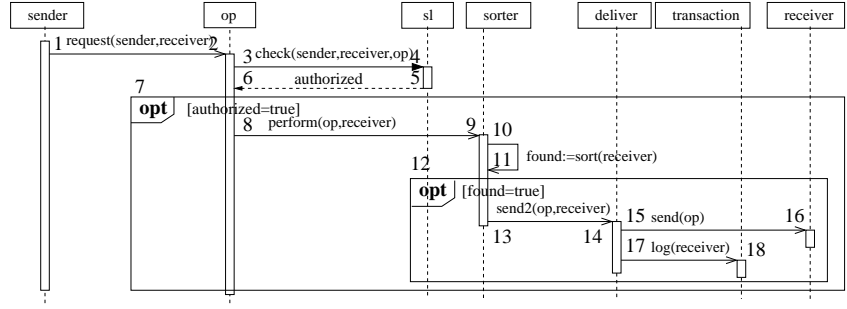}}
\end{center}
\caption{An SD for an application that enforces Mandatory Access
  Control}
\label{fig:mac2}
\end{figure*}

\begin{example}\hspace{1pc}  \label{ex:mac2}
Consider the SD in Fig.~\ref{fig:mac2} for email communication where
fragments and events are labelled.  In particular, the outer \textit{opt}
fragment is labelled $7$. The sending and receiving events for a
message in the SD are labelled with two consecutive numbers. Let $e_i$
abbreviate the event labelled $i$ as $e_i$. For instance, $e_1$
abbreviates $(!,request(sender,receiver), sender,op)$. Then
the SD is expressed as \( D_{App} = block( \{ 1..7\},  \{i\mapsto
e_i\mid 1\leq i \leq 6\} \cup\{7\mapsto f_7\}, \strict_0)\) where
$\strict_0=\{\langle i,{i+1}\rangle\mid 1\leq i\leq 6\}$ and $f_7=
opt(authoried=true, block(\{8..12\}, \{i\mapsto e_i \mid 8\leq i\leq
11\}\cup\{12\mapsto f_{12}\}, \strict_1))$ with $\strict_1 =\{\langle i,{i+1}\rangle\mid 8\leq
i\leq 11 \}$,
$f_{12}=block(\{13..18\}, \{i\mapsto e_i\mid 13\leq i\leq 18\},\{\langle i,i+1\rangle \mid 13 \leq i\leq 15\}\cup\{\langle 15,17\rangle,\langle 17, 18\rangle\})$.
\end{example}

\section{Semantics} \label{sec:semantics}
This section presents the semantic domain and the semantic equations
for the trace semantics.  \add{We first introduce auxiliary notations
  and operations used in the construction of the domain and the
  definition of the semantic equations.}

Let $\Sigma$ be an alphabet. $\Sigma^\ast$ denotes the set of all
strings over $\Sigma$. The empty string is denoted $\epsilon$.  A
language $L$ over $\Sigma$ is a set of strings over $\Sigma$.  The
Kleene closure of $L$ is denoted $L^\ast$.  Let $\omega\in\Sigma$.
The length of $\omega$ is denoted $|\omega|$. The string $\omega$
may be thought of as a function from $\{0..|\omega|-1\}$ to
$\Sigma$. The $i$-th element in $\omega$ is written as
$\omega(i)$.  The interleave of two strings is the set of strings
obtained by interleaving the two strings in all possible ways. Let
$x,y\in\Sigma$ and $\mu,\nu\in\Sigma^\ast$. The following
definition of the interleave operator $\interleave$ is due to
\cite{Storrle03}.
\begin{eqnarray*}
\epsilon \interleave \mu &= &
\mu\interleave \epsilon = \mu\\
x\mu\interleave y\nu &=& \{x\}\bullet (\mu\interleave y\nu) \cup
\{y\}\bullet (x\mu\interleave v)
\end{eqnarray*}
where $\bullet$ is the language concatenation operator.

Let $\oplus$ be a binary operation on domain $S$. Then
$\oplus^\sharp$ denotes this binary operation
 on $\wp(S)$ defined as follows.
\[X~ \oplus^\sharp~ Y = \{x~\oplus~y \mid x\in X \wedge y\in Y\} \]
For instance, $\cap^\sharp,\cup^\sharp$ and $\bullet^\sharp$ are
respectively pair wise set intersection, set union and language
concatenation.

{
\begin{eqnarray*}
 \mathcal{M}_1\cap^\sharp \mathcal{M}_2 &=& \{ \mathcal{C}_1\cap \mathcal{C}_2 \mid
\mathcal{C}_1\in \mathcal{M}_1 \wedge \mathcal{C}_2\in\mathcal{M}_2 \}\\
 \mathcal{M}_1\cup^\sharp \mathcal{M}_2 &=& \{ \mathcal{C}_1\cup \mathcal{C}_2 \mid
\mathcal{C}_1\in \mathcal{M}_1 \wedge \mathcal{C}_2\in\mathcal{M}_2 \}\\
 \mathcal{M}_1\bullet^\sharp \mathcal{M}_2 &=& \{ \mathcal{C}_1\bullet \mathcal{C}_2 \mid
\mathcal{C}_1\in \mathcal{M}_1 \wedge \mathcal{C}_2\in\mathcal{M}_2 \}
\end{eqnarray*}
}

\add{A rewriting relation $\Rightarrow$ on a set $A$ is a binary
  relation on $A$. An element $a\in A$ is a normal form if there is no
  $a'\in A$ such that $a\Rightarrow a'$. A rewriting relation
  $\Rightarrow$ is called finitely terminating iff it has no infinite
  descending chain $a_0\Rightarrow a_1 \Rightarrow a_2\cdots$. It is
  called confluent if, for each $x,u,w\in A$ such that
  $x\Rightarrow^\ast u$ and $x\Rightarrow^\ast w$, there is a $z$
  such that $u\Rightarrow^\ast z$ and $w \Rightarrow^\ast
  z$. $\Rightarrow$ is called convergent if it is both confluent and
  finitely terminating. If $\Rightarrow$ is convergent, then for each
  $a$ there is a unique normal form $a'$, denoted $a_{\Rightarrow}$,
  such that $a\Rightarrow^\ast a'$~\cite{BaaderT98}.}

Let $\dom(f)$ be the domain of a function $f$ and
$image(f)=\{f(x)\mid x\in \dom(f)\}$ the image of $f$.

\subsection{Semantic Domain}
 An SD is a partial specification of required and prohibited
 behaviors of an application. This paper is concerned only with
 required behaviors. Consider this simple SD.
\begin{center}
\scalebox{0.8}{\includegraphics{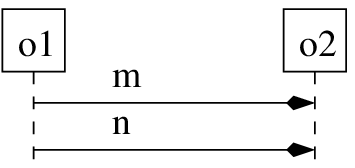}}
\end{center}
{ The SD has four events $!m$,$?m$, $!n$ and $?n$ and its
  strict sequencing order is $\strict=\{\langle !m,?m\rangle, \langle
  !n,?n\rangle, \langle !m,!n\rangle, \langle?m,?n\rangle \}$.} An
implementation that produces the trace $!m?m!n?n$ satisfies this
specification; another implementation that produces the trace
$!m!n?m?n$ also satisfies the specification. Thus, the SD specifies
two alternative minimum obligations $\mathcal{O}_1=\{ !m ?m !n ?n\}$
and $\mathcal{O}_2=\{ !m ?m !n ?n\}$. Of course, an implementation
that non-deterministically produces one of the two traces also
satisfies the specification. However, the obligation
$\mathcal{O}_1\cup\mathcal{O}_2$ is redundant since it includes
$\mathcal{O}_1$ and $\mathcal{O}_2$ as proper subsets and hence not
minimum.  An obligation may contain more than one traces. Once an
obligation is chosen, all traces in the obligation are required in
that for each trace $t$ in the obligation, there is an interaction
that produces $t$. For instance, $alt(v=ok,!m, !n)$ has one obligation
with two traces $\{(v=ok) !m, (v\neq ok) !n\}$. A condition such as
$(v=ok)$ is a guard for the rest of the trace, meaning that the rest
is exhibited only if the condition evaluates to true. The above
obligation requires an implementation to produce $!m$ if $(v=ok)$ is
true when it runs and to produce $!n$ if $(v\neq ok)$ is true.

A critical fragment requires that there is not any intervening event
between two consecutive events in the region. For instance, a critical
fragment in the specification of a telephone service may specify that
after receiving a 911 call from the user, the operator must forward
the call to the emergency service without any
interruption. {Another example is  the
  specification for a home security system. It  may specify that after
  receiving an abnormal response from the sensor, the alarm cell must
  set off the alarm device and alert the security agency and these
  messages must occur as an uninterrupted sequence.}  We wrap a
sub-trace from a critical fragment and treat it as a single token.

 A trace is a sequence of tokens which are either events, guard
 conditions or critical segments $\region{\sigma}$ where $\sigma$ is a
 sequence of events and guard conditions. A critical segment
 $\region{\sigma}$ protects the sub-trace $\sigma$ from
 interference. Occurring in a trace, $\region{\sigma}$ will be treated
 as atomic when the trace is combined with other traces through
 interleaving and weak sequencing. The domains of tokens and traces
 are respectively
 \begin{eqnarray*}
\tokens &=& \events \cup \constraints \cup
\region{(\events \cup \constraints)^\ast}\\
\traces & = & \tokens^\ast
\end{eqnarray*}

Consider two required traces $c!m$ and $\neg_{c}(c) !m$. Then message
$m$ is always sent since it is always the case that either $c$ or
$\neg_{c}(c)$ holds. Occurring in an obligation, they represent an
unnecessary decision point. \change{Define $\mathcal{O}\cup \{\alpha
  c\beta,\alpha c'\beta\} \stackrel{fold}{\leadsto}
  \mathcal{O}\cup\{\alpha\beta\}$ if $c\vee_{c} c'\models_{c}
  true$. For any $\mathcal{O}$, there is a unique $\mathcal{O}'$
  denoted $\mathit{fold}(\mathcal{O})$ such that $\mathcal{O}
  \stackrel{fold}{\leadsto}^\ast \mathcal{O}'$ and
  $\mathcal{O}'\not\stackrel{fold}{\leadsto}$.  }{ Define $\multimap$
  by $\mathcal{O}\cup \{\alpha c\beta,\alpha c'\beta\} \multimap
  \mathcal{O}\cup\{\alpha\beta\}$ if $c\vee_{c} c'\models_{c}
  true$. Then $\multimap$ is a convergent
  rewriting relation on obligations. So, function
  $\mathit{fold}(\mathcal{O})=\mathcal{O}_{\multimap}$ is well
  defined.} The function is lifted to sets of obligations
as $\mathit{fold}(\mathcal{M}) = \{\mathit{fold}(\mathcal{O}) \mid
\mathcal{O}\in\mathcal{M}\}$.  Define
\[ \downarrow \mathcal{M}  = \{ \mathcal{O} \in \mathcal{M} \mid \neg \exists \mathcal{O}'\in\mathcal{M}. (\mathcal{O}'\subset \mathcal{O})\} \]
The operation $\downarrow$ removes redundancy from its argument.
The semantic domain is
\[
\mathbbm{Sem} = \{ \mathcal{M} \in \wp(\wp(\traces)) \mid
\mathcal{M}=\downarrow\mathcal{M} \wedge
\mathit{fold}(\mathcal{M})= \mathcal{M} \}
\]

\subsection{Semantic function}
The semantics of an SD $D$ is denoted $\sem{D}$. It is defined as the
least solution to a system of semantic equations.

\subsubsection{Observable and unobservable events.}
We are now ready for defining semantic equations. Observable and
unobservable events have obvious semantics:
\begin{eqnarray*} \sem{e} &=& \{\{ e\}\}\\
\sem{\tau} &=&\{\{\epsilon\}\}
\end{eqnarray*}
where $\epsilon$ is the empty trace.

\subsubsection{Strict fragments.} The concatenation of an
obligation $\mathcal{O}_1$ in $\sem{D_1}$ and an obligation
$\mathcal{O}_2$ in $\sem{D_2}$ \change{is}{gives rise to} an
obligation $\mathcal{O}$ in $\sem{strict(D_1,D_2)}$.
\[ \sem{strict(D_1,D_2)} =
\downarrow \mathit{fold}(\sem{D_1}\bullet^\sharp \sem{D_2})\]
\add{where $\mathit{fold}$ is applied to eliminate unnecessary
  decision points and $\downarrow$ to remove redundant
  obligations. These functions are also applied in semantic functions
  for other kinds of fragment.}

\subsubsection{Critical fragments.}
The semantics of $critical(D)$ is defined by unwrapping all critical
segments in traces of $D$ and wrapping the result in
$\region{\cdot}$.  \change{ Define
  \(\alpha\region{\sigma}\beta\curvearrowright
  \alpha\sigma\beta\). The relation $\curvearrowright$ exposes a
  sub-trace protected by a critical segment.  For any $\sigma$, there
  is a unique $\eta$ such that $\sigma\curvearrowright^\ast\eta$ and
  $\eta\not\curvearrowright\omega$ for any $\omega$. Let ${\it
    unwrap}$ be the function that maps $\sigma$ to $\eta$ and define} { Define
  $\curvearrowright$ by \(\alpha\region{\sigma}\beta\curvearrowright
  \alpha\sigma\beta\). A rewriting step with $\curvearrowright$ exposes a
  sub-trace protected by a critical segment. Since  $\curvearrowright$  is convergent,
  $\mathit{unwrap}(\sigma)=\sigma_{\curvearrowright}$ is a well defined function. Define}
 $\mathit{wrap}(\sigma) =\region{\sigma}$.  The function
$\mathit{unwrap}$ is lifted to sets of sets as
$\mathit{unwrap}(\mathcal{M}) = \{ \{ \mathit{unwrap}(\omega)\mid
\omega\in\mathcal{\mathcal{O}}\} \mid
\mathcal{\mathcal{O}}\in\mathcal{M}\}$.  Lift $\mathit{wrap}$ in the
same way. The semantics of $critical(D)$ is defined \[
\sem{critical(D)} =
\mathit{wrap}(\downarrow\mathit{fold}(\mathit{unwrap}(\sem{D})))\] For
instance, $\sem{critical(strict(e,f))}= \{\{\region{ef}\}\}$ where
$e,f\in\events$.

\subsubsection{Alt fragments.}
The semantics of $alt(c,D_1,D_2)$ is obtained by pre-pending $c$
to traces of $D_1$ and $\neg_{c}(c)$ to those of $D_2$: \[
\sem{alt(c,D_1,D_2)} = \downarrow
\mathit{fold}(\{\{c\}\}\bullet^\sharp \sem{D_1} \cup^\sharp
\{\{\neg_{c}(c)\}\}\bullet^\sharp \sem{D_2})\]
{Let
$e,f,g\in\events$ and $c\in\constraints$. Then}
 $\sem{alt(c,e,e)}=\{\{e\}\}$ and\linebreak
$\sem{alt(c,strict(e,f),g)} =\{\{cef,c'g\}\}$ where $c'=\neg_{c}(c)$.

\subsubsection{Opt fragments.} The semantics of $opt(c,D)$ is
obtained similarly: \[ \sem{opt(c,D)} =
\downarrow\mathit{fold}(\{\{c\}\}\bullet^\sharp \sem{D} \cup^\sharp
\{\{\neg_{c}(c)\}\})\] For instance,
$\sem{opt(c,\tau)}=\{\{\epsilon\}\}$.

\subsubsection{Par fragments.}
Consider parallel interleave $par(D_1,D_2)$ of two sub-diagrams.
Let $\mathcal{O}_1$ be an obligation of $D_1$ and $\mathcal{O}_2$
of $D_2$.  Parallel interleaving  produces a set of alternative
obligations from $\mathcal{O}_1$ and $\mathcal{O}_2$. Define
\[ \mathcal{O}_1 \hat{\interleave} \mathcal{O}_2 =
\{ \mathcal{O} \mid \forall \sigma_1\in \mathcal{O}_1.\sigma_2\in \forall \mathcal{O}_2.\exists \sigma\in \mathcal{O}.
(\sigma\in \sigma_1\interleave \sigma_2) \}
\]
$\mathcal{O}_1 \hat{\interleave} \mathcal{O}_2$ may have redundant
obligations. Put $\mathcal{O}_1=\{e_1\}$ and
$\mathcal{O}_2=\{e_2\}$. Then $\mathcal{O}_1
\hat{\interleave}\mathcal{O}_2= \{\{e_1e_2\},\{e_2e_1\},
\{e_1e_2,e_2e_1\} \}$. The obligation $\{e_1e_2,e_2e_1\} $ is
redundant. The meaning of $par(D_1,D_2)$ is defined
\[\sem{par(D_1,D_2)} = \sem{D_1} ~\interleave^\flat~\sem{D_2}\]
where
\[ \mathcal{M}_1 \interleave^\flat  \mathcal{M}_2 =
\downarrow \left(\bigcup_{\mathcal{O}_1\in \mathcal{M}_1,
\mathcal{O}_2\in \mathcal{M}_2} \mathcal{O}_1\hat{\interleave}
\mathcal{O}_2\right)
\]
\begin{example}\hspace{1pc}  Let $c\in\constraints$ and $f,g,h\in\events$.
Then
\[
\begin{array}{l}
 \sem{par(alt(c,f,g),h)} = \{\{cf,c'g\}\} ~\interleave^\flat~ \{\{h\}\}\\
~~= \left\{ \begin{array}{c}
    \{hcf,hc'g\}, \{hcf,c'hg\},\{hcf,c'gh\}, \\
   \{chf,hc'g\},  \{chf,c'hg\}, \{chf,c'gh\}, \\
   \{cfh,hc'g\}, \{cfh,c'hg\},\{cfh,c'gh\}
\end{array}\right\}
\end{array}\] where $c'=\neg_{c}(c)$. \end{example}

\subsubsection{Block fragments.}
To enforce sequencing orders, we tag tokens in a trace generated from
a fragment. \change{Define}{Function $\tag$ labels each token in a
  trace by a given label}: $\tag(\epsilon,\ell) = \epsilon$ and
$\tag(t\cdot\sigma,\ell) =\langle
t,\ell\rangle\cdot\tag(\sigma,\ell)$.  Function $\untag$ does the
opposite and is defined $\untag(\epsilon) = \epsilon$ and
$\untag(\langle
t,\ell\rangle\cdot\hat{\sigma})=t\cdot\untag(\hat{\sigma})$.  $\tag$
and ${untag}$ are extended to sets of sets in the same way as
$\mathit{unwrap}$.  Function $\lifelines$ maps a token to the set of
the lifelines associated with the token. A sending event is associated
with the sender, a receiving event with the receiver and a critical
segment with all the lifelines associated with the events in the
critical segment.  $\lifelines$ is defined by $\lifelines(\langle
!,N(P),S,R\rangle)=\{S\}$, $\lifelines(\langle
?,N(P),S,R\rangle)=\{R\}$, $\lifelines(\region{\sigma})=\bigcup_{i\in
  dom(\sigma)}\lifelines(\sigma(i))$ and
$\lifelines(c)=\emptyset$. Let $\mathit{lb}$ be the function that
returns the label of a tagged token.  Then $\mathit{lb}(\langle
t,\ell\rangle)=\ell$.  Relation $\sim$ relates two tagged tokens iff
they share lifelines: $\langle t_1,\ell_1\rangle \sim \langle
t_2,\ell_2\rangle$ iff $\lifelines(t_1) \cap\lifelines(t_2) \neq
\emptyset$. Let $\taggedtraces = (\tokens\times
\labels)^\ast$\add[LL]{ be the set of tagged traces}. The set of
traces of tagged tokens satisfying a strict sequencing order $\strict$
is denoted $\mathit{st}(\strict)$.
\[\mathit{st}(\strict) =
\left\{\hat{\sigma}\in \taggedtraces~
\begin{array}{|l} \forall 0 \leq i,j < |\hat{\sigma}|.\\
~~((\mathit{lb}(\hat{\sigma}(i)) \strict^\ast
\mathit{lb}(\hat{\sigma}(j)) \Rightarrow (i\leq j))\end{array} \right\}
\]
 The semantics of block fragments is
defined as
\[
\sem{block(L,\iota, \strict)} = \untag(\{\mathit{st}(\strict)\}
\cap^\sharp (\interleave^\flat_{\ell\in L}
\mathit{tag}(\sem{\iota(\ell)},\ell)))
\]
\add{Traces from immediate sub-fragments of $block(L,\iota, \strict)$
  are first interleaved in all possible ways and then those traces are
  removed that violate the strict sequencing order $\strict$.  The
  labels that are used to tag tokens do not occur in the resulting
  semantics; they are only used in enforcing the strict sequencing
  order $\strict$.}

\begin{example}\hspace{1pc}  \label{ex:running:semantics}
Continue with Example~\ref{ex:running:syntax}. We have\linebreak
 $\sem{D_a}=
\downarrow\mathit{fold}(\{\{[OK=true]\}\}\bullet^\sharp [\!\![block(\{7,8\},
  \{7\mapsto e_7, 8 \mapsto$ $ e_8\},\{\langle
  7,8\rangle\})]\!\!] \cup^\sharp \{\{[OK=false]\}\}) = \{\{[OK=true]e_7e_8,$ $ [OK=false]\}\}
$ and $\sem{\mathit{Login}} = \{\{r_1,r_2\},\{r_1',r_2'\}\}$ where
$r_1,r_2,r_1'$ and $r_2'$ are given in Example~\ref{ex:behavior}.
\end{example}

\subsubsection{Seq fragments.} The interaction operator
\textit{seq} combines traces from component SDs via weak
sequencing. The semantics of $\mathit{seq}(D_1,D_2)$ is obtained as
follows.  Every token in each trace in $\sem{D_1}$ is tagged with 1
and every token in each trace of $\sem{D_2}$ is tagged with 2.  Tagged
traces are then interleaved as in the semantics of
$\mathit{par}(D_1,D_2)$. Then any tagged trace that violates weak
sequencing order imposed by \textit{seq} is removed. The set of tagged
traces that satisfy the weak sequencing order is
\[
\taggedtraces_{seq}  =
\left\{\hat{\sigma}\in \taggedtraces
\begin{array}{|l}
\forall 0\leq i,j < |\hat{\sigma} |. \\
\left( \begin{array}{c}
  (\mathit{lb}(\hat{\sigma}(i)) =1 \\ \wedge \\
\mathit{lb}(\hat{\sigma}(j)) =2 \\ \wedge \\  (\hat{\sigma}[i] \sim
\hat{\sigma}[j]) \end{array}\right) \Rightarrow (i<j)) \end{array}\right\}
\]
The semantics of weak sequencing fragments is   defined
\[
\sem{seq(D_1,D_2)} = \sem{D_1}\merge^\flat \sem{D_2}
\] where
\[ \mathcal{M} \merge^\flat \mathcal{N} =
 \mathit{untag}( \{\taggedtraces_{seq}\} \cap^\sharp ( \mathit{tag}(\mathcal{M},1)
\interleave^\flat
\mathit{tag}(\mathcal{N},2)))
\]

\begin{example}\hspace{1pc}  \label{ex:sem}
Let $o_1,o_2$ be lifelines,
 $f_1= (!,m, o_1,o_2)$, $f_2= (?,m,o_1,o_2)$,
 $f_3= (!,n,o_1,o_2)$ and  $f_4= (?,n,o_1,o_2)$. Put
$D_1= strict(f_1,f_2)$ and $D_2=strict(f_3,f_4)$. Then
{\setlength\arraycolsep{0.1em}
\begin{eqnarray*}
\sem{D_1} &=& \{\{f_1f_2\}\}\\
\sem{D_2} &=& \{\{f_3f_4\}\}\\
\sem{strict(D_1,D_2)} &=& \{\{f_1f_2f_3f_4\}\}\\
\sem{seq(D_1,D_2)} &=& \{\{f_1f_2f_3f_4\},\{f_1f_3f_2f_4\}\}\\
\sem{par(D_1,D_2)} &=& \{\{f_1f_2f_3f_4\},\{f_1f_3f_2f_4\},
\{f_1f_3f_4f_2\}, \\ && ~\{f_3f_4f_1f_2\},\{f_3f_1f_4f_2\},
\{f_3f_1f_2f_4f_4\} \}
\end{eqnarray*}
}
 Let $\iota(i)=D_i$ for $i=1,2$. The $\sem{block(\{1,2\},\iota,
\{\langle 1,2\rangle\})} =\sem{strict(D_1,D_2)}$,
 and
$\sem{block(\{1,2\},\iota,\emptyset)}
=\sem{par(D_1,D_2)}$.

 \end{example}

\subsubsection{Loop fragments.}
The UML standard stipulates that traces from consecutive runs of the
loop body are combined via weak sequencing: $\sem{loop(c,D)}$ is the
limit of this series: $X_0 = \{\{\neg_{c}(c)\}\}$ and $X_{i+1} =
(\{\{c\}\}\bullet^\sharp (\sem{D}\merge^\flat X_i))
\cup^\sharp\{\{\neg_{c}(c)\}\}$.  {An alternative definition
  would be
\[ \sem{loop(c,D)} = (\{\{c\}\}\bullet^\sharp
\sem{seq(D,loop(c,D))})\cup^\sharp\{\{\neg_{c}(c)\}\}
\]}

\subsubsection{Properties of semantics.}
The abstract syntax requires that the $block$ fragment has at least
one immediate sub-fragment. As a consequence, a sequence diagram
specifies at least one obligation.

\begin{lemma} \label{lm:one} Let $D\in\diagrams$. Then
$\exists\mathcal{O}\in\sem{D}.(\mathcal{O}\neq\emptyset)$.
\end{lemma}

Let $\events(D)$ be the set of observable events occurring in $D$.
\begin{lemma} If $\events(D)=\emptyset$ then
$\sem{D}=\{\{\epsilon\}\}=\sem{\tau}$. \label{lm:two}
\end{lemma}

We adapt the concept of a context from term writing.  A context is an
SD with one of its fragments replaced by a special symbol
$\mathbbm{x}$. For instance, $seq(\mathbbm{x},e)$ with $e\in\events$ is
a context. Let $D$ be an SD and $C$ a context. The embedding of $D$
into $C$, denoted $C[D]$ is the SD obtained from replacing
$\mathbbm{x}$ with $D$.  Two SDs are called equivalent if they have
the same meaning.  The following proposition shows that the semantics
possesses substitutivity.  Substitutivity is a desirable property
since it allows any fragment in an SD to be replaced with a
semantically equivalent fragment.

\begin{proposition} \label{prop} Let $C$ be a context and $D_1,D_2\in\diagrams$.
If $\sem{D_1}=\sem{D_2}$ then $\sem{C[D_1]}= \sem{C[D_2]}$.
\end{proposition}

\section{Semantics based Conformance} \label{sec:preservation}
In this section, we make precise of the notion of conformance.
\comment
{Non-determinism in an SD leads to multitude of
alternative obligations.  A system implementing an SD $D$
must fulfill at least one of the obligations.  When an SD
is refined, the designer eliminates non-determinism and makes the
specification more defined. In doing so, he must ensure that any
implementation of the refined model is also an implementation of the
original model.
}
There are a number of issues to consider in reasoning about SD
conformance.  One issue is renaming of lifelines, messages and system
variables.  When reusing an SD, the designer embeds it into a
context. In doing so, the designer may need to change the names of
lifelines and messages either for better conveying his intention or
for avoiding name conflicts. Another issue is the introduction of new
lifelines, messages and system variables which are unobservable in the
original SD. The values that the unobservable system variables take
affect the behavior of the specified system.  Yet another issue is the
use of guard conditions in fragment combination operators.  The
conformance relation we shall define is parameterized by a mapping
\remove[LL]{that renames lifelines and assigning values to system
  variables} $\rho:\names\mapsto\names$ and a set of
  events $\mathcal{U} \subset \events$.  \add[LL]{The mapping $\rho$ is
    called a substitution and it maps new names of lifelines, messages
    and system variables to their old names and assigns values to
    newly introduced system variables. Application of a substitution
    $\rho$ to a syntactic object $o$, denoted $\rho(o)$, is obtained
    from substituting $\rho(n)$ for each occurrence of each name $n$
    in $o$.}  The latter induces a hiding function
  $hide_{\mathcal{U}}$ on $\diagrams$. $hide_{\mathcal{U}}(D)$ is the
  SD obtained from $D$ by replacing all occurrences of $e$ with $\tau$
  for each $e\in\mathcal{U}$.

\subsection{Trace Simulation Relation}
We first define a simulation relation between traces that take into
account the use of guard conditions.

\begin{definition} \add{Let  $c_1,c_2\in\constraints$, $e_1,e_2\in\events$ and
 $\alpha,\beta,\gamma\in\traces$.}
 The trace simulation relation $\ltimes$ is defined \add[LL]{inductively} as follows.
\begin{itemize}
\item $c_1 \ltimes  c_2$ \change[LL]{iff}{if} $c_2 \models_{c} c_1$.
\item $e_1 \ltimes e_2$ \change[LL]{iff}{if} $e_1=e_2$.

\item $\region{\alpha}\ltimes \region{\gamma}$ \change[LL]{iff}{if}
  $\alpha\ltimes \gamma$,

\item $\alpha\ltimes \gamma$ \change[LL]{iff}{if} there are a trace
  $\beta$ such that $\alpha\curvearrowright^\ast\beta$ and a strictly
  increasing function $\eta:\dom(\beta)\mapsto
  dom(\gamma)$ such that

\begin{itemize}
 \item[(1)] for any $i\in\dom(\beta)$, $\beta(i)\ltimes \gamma(\eta(i))$; and

\item[(2)] for $j\in\dom(\gamma)$, if $j\not\in\mathit{image}(\eta)$ then
$\gamma(j)\in\constraints$.

\end{itemize}
\end{itemize} \label{df:trace-conform}
\end{definition}

Some explanations are in order.  A critical segment can only be
simulated by a critical segment.  The condition
$\alpha\curvearrowright^\ast\beta$ allows events in protected
sub-traces to be used to simulate events in $\gamma$ by breaking up
zero or more occurrences of $\region{\cdot}$. Note that $\beta$ may be
$\alpha$ itself.  The strict monotonicity of $\eta$ ensures that
different events in $\gamma$ are simulated by different events in
$\beta$.  The condition (2) ensures that if the events in $\gamma$
occur then the events in $\beta$ occur too. The condition (1)
guarantees that each event in $\gamma$ is simulated by an event in
$\beta$.

\begin{lemma} \label{lm:aux1} If $\alpha_1 \ltimes  \alpha_2$ and
$\alpha_2\curvearrowright \beta_2$ then there is a $\beta_1$ such that
$\alpha_1\curvearrowright \beta_1$ and  $\beta_1 \ltimes  \beta_2$.
\end{lemma}

The following is the consequence of the reflexivity and transitivity
of $\models_{c}$.
\begin{lemma} \label{lm:aux2}
$\ltimes $ is reflexive and transitive.
\end{lemma}

\begin{example}\hspace{1pc}
Let $e_1,e_2,e_3$ be different events and $c$ a guard condition.  Then
$e_1\cdot e_3\ltimes e_1\cdot e_3$ and $\region{e_1\cdot e_3}\ltimes
\region{e_1\cdot c\cdot e_2}$. But, $\region{e_1\cdot e_2\cdot
  e_3}\ltimes\region{e_1\cdot e_3}$ does not hold since $e_2$ is an
event and it is between $e_1$ and $e_3$.  Nor does $c\cdot e_1\ltimes
e_1$ hold since there is no guarantee that the constraint $c$ is
satisfied.
\end{example}

\subsection{Refinement Relation}
We now introduce a special case of conformance called refinement.  An
SD specifies a number of alternative obligations and an implementation
may choose to realize any of them. An SD $D_1$ refines another SD
$D_2$ if any implementation of $D_1$ is also an implementation of
$D_2$. Formally,
\begin{definition}
Let $D_1,D_2\in\diagrams$. $D_1$ is said to refine $D_2$, denoted
$D_1\succeq D_2$, if \( \forall \mathcal{O}_1\in\sem{D_1}.\exists
\mathcal{O}_2\in\sem{D_2}.\forall t_2\in \mathcal{O}_2.\exists t_1\in
\mathcal{O}_1. (t_1\ltimes t_2) \).
\end{definition}

The following lemmas follow from definitions of $hide_{\mathcal{U}}$,
$\sem{\cdot}$ and $\ltimes$. They state that both hiding and
substitution preserve refinement relation between SDs.

\begin{lemma} Let $D_1$ and $D_2$  be SDs. If
 \( {D_1} \succeq D_2 \) then \( {hide_{\mathcal{U}}(D_1)} \succeq {
   hide_{\mathcal{U}}(D_2)} \) for any $\mathcal{U}\subseteq\events$.
\label{remark1}
\end{lemma}

\begin{lemma} Let $D_1$ and $D_2$  be SDs.
If
 \( {D_1} \succeq D_2 \) then \( {\rho(D_1)} \succeq {\rho} \)
for any substitution
 $\rho:\names\mapsto\names$.
\label{remark2}
\end{lemma}

\subsection{Conformance Relation}

 We are now ready to define the conformance relation between SDs.
 \change[LL]{If we change $D_2$ into $D_1$ by adding new events in
   $\mathcal{U}$, changing names of lifelines, messages and system
   variables in $D_2$ and introducing new system variables, we need to
   make sure that any implementation of
   $\rho(hide_{\mathcal{U}}(D_1))$ is also an implementation of $D_2$
   where $\rho$ is a mapping that renames lifelines and assigning
   values to new system variables.}  {If we change $D_2$ to $D_1$, we
   need to make sure that $\rho(hide_{\mathcal{U}}(D_1))$ refines
   $D_2$ where $\mathcal{U}$ is the set of newly introduced events,
   $\rho$ is a substitution that reverses name changing and assigns
   values to new system variables.}  It is also necessary to make sure
 that events in $\mathcal{U}$ are not those that are used to simulate
 events in $\events(\rho(D_2))$.

\begin{definition}\label{df:sd-conform}
Let $D_1,D_2\in\diagrams$,  $\rho:\names\mapsto\names$ and $\mathcal{U}\subseteq\events(D_1)$.
 We say that $D_1$ conforms to $D_2$ with respect to $\rho$ and $\mathcal{U}$, denoted $D_1
\conformsto_{\rho,\mathcal{U}}  D_2$ iff
\begin{enumerate}
\item $\rho(\mathcal{U}) \cap \events(D_2) =\emptyset$, and
\item
$\rho(hide_\mathcal{U}(D_1))\succeq D_2$, i.e., $\rho(hide_\mathcal{U}(D_1))$ refines $D_2$.
\end{enumerate}

 We say that $D_1$ conforms to $D_2$, denoted $D_1 \conformsto D_2$ iff $D_1
 \conformsto_{\rho,\mathcal{U}} D_2$ for some
 $\rho:\names\mapsto\names$ and some $\mathcal{U}\subseteq\events(D_1)$.

\end{definition}
Note that refinement is a special case of conformance in which
$\mathcal{U}=\emptyset$ and $\rho$ is the identity function. In
other words, $D_1$ conforms to $D_2$ whenever $D_1$ refines $D_2$.

\begin{example}\hspace{1pc}  Continue with Example~\ref{ex:sem}.
Let $\rho$ be the identity function and $\mathcal{U}=\emptyset$.
It can be verified that $strict(D_1,D_2) \conformsto seq(D_1,D_2)$
and\linebreak $seq(D_1,D_2) \conformsto par(D_1,D_2)$.
\end{example}

\begin{theorem} The conformance relation $\conformsto$ is reflexive and transitive, i.e.,
\begin{enumerate}
\item
$D~\conformsto ~ D$ for any $D\in\diagrams$;
\item
 if $D_1
  ~\conformsto ~ D_2$ and $D_2
  ~\conformsto ~ D_3$ then $D_1
  ~\conformsto~  D_3$ for any
  $D_1,D_2,D_3\in\diagrams$.
\end{enumerate}
\label{th:1}
\end{theorem}

\begin{example}\hspace{1pc}
This example shows that SD Login2 conforms to SD Login. Let $f_i$ denote
the event that is labelled ${i}$.  Then $\mathit{Login2}=
block(\{b,5,6,11,12,c\},\\ \{b\mapsto D_b, 5\mapsto f_5, 6\mapsto f_6,
11\mapsto f_{11}, $ $ 12\mapsto f_{12}, c\mapsto D_c\}, \{\langle
b,5\rangle, \langle 5,6\rangle,\langle 6,11\rangle,$ $ \langle
11,12\rangle,$ $ \langle 12,c\rangle\}) $ with
$D_b=strict(block(\{1,2\},\{1\mapsto f_1,2\mapsto f_2\},\{\langle
1,2\rangle\}),$ $strict(block( $ $\{3,4\},$ $\{3\mapsto f_3,4\mapsto
f_4\},\{\langle 3,4\rangle\}),$ $ block(\{9,10\},\{9\mapsto f_9,10\mapsto
f_{10}\},\{\langle 9,10\rangle\})))$ and $D_c=opt(pOK=true\wedge
kOK=true,block(\{7,8\},\{7\mapsto f_7,8\mapsto f_8\},\{\langle
7,8\rangle\}))$. Let
$\mathcal{U}= \{f_9,f_{10},f_{11},f_{12}\}$ and
\[\rho=\left\{ \begin{array}{c}
{customer}\mapsto{user}, {brokerage}\mapsto{server},\\
{acc}\mapsto{id},  {pin}\mapsto{pwd}, {chkP}\mapsto{chk},\\
{pOK}\mapsto{OK},  {trade}\mapsto{cmd},{kOK}\mapsto true
\end{array}\right\}\]
Then $\rho(f_i)=e_i$ for $1\leq i \leq 8$ and
\begin{eqnarray*}
\lefteqn{\rho(hide_{\mathcal{U}}(D_b)) = }\\
&&\begin{array}[t]{l} strict(block(\{1,2\},\{1\mapsto e_1,2\mapsto e_2\},\{\langle
1,2\rangle\}),\\  ~~~~~strict(block(\{3,4\},\{3\mapsto e_3,4\mapsto
e_4\},\{\langle 3,4\rangle\}),\\ ~~~~~~~~~~block(\{9,10\},\{9\mapsto \tau,10\mapsto\tau\},\{\langle 9,10\rangle\}))) \end{array}
\end{eqnarray*}
By the definition of $\sem{\cdot}$,
{\setlength\arraycolsep{0.1em}
\begin{eqnarray*}
\sem{block(\{9,10\},\{9\mapsto \tau,10\mapsto\tau\},\{\langle
  9,10\rangle\})} &=& \{\{\epsilon\}\}\\
\sem{block(\{1,2\},\{1\mapsto e_1,2\mapsto e_2\},\{\langle
  1,2\rangle\})} &=& \{\{e_1e_2\}\}\\
\sem{block(\{3,4\}, \{3\mapsto
  e_3,4\mapsto e_4\},\{\langle 3,4\rangle\})}&=&\{\{e_3e_4\}\}\\
\sem{\rho(hide_{\mathcal{U}}(D_b))} &=&
\{\{e_1e_2e_3e_4\}\}\end{eqnarray*} } and \(
\rho(hide_{\mathcal{U}}(D_c)) =
opt(OK=true,block(\{7,8\},\{7\mapsto e_7, 8\mapsto e_8\},\{\langle
7,8\rangle\})) \) after the tautology $true=true$ is removed from
the guard condition. Thus,
\[ \sem{\rho(hide_{\mathcal{U}}(D_c)) } =
\{\{[OK=true]e_7e_8,[OK=false]\}\}
\]
Finally, let $Login'= \rho(hide_{\mathcal{U}}(\mathit{Login2}))$,
$D_b'= \rho(hide_{\mathcal{U}}(D_b))$ and $D_c'=
\rho(hide_{\mathcal{U}}(D_c))$. Then $Login' =
block(\{b,5,6,11,12, c\}, \{b\mapsto D_b', 5\mapsto e_5, 6\mapsto e_6,
11\mapsto \tau, 12\mapsto \tau, c\mapsto D_c'\}, \{\langle b,5\rangle,
\langle 5,6\rangle,\langle 6,11\rangle,\langle 11,12\rangle, \langle
12,c\rangle\}) $. Since the strict sequencing order in SD $Login'$ is
total, traces from its components are combined using string
concatenation and
\begin{eqnarray*} \sem{Login'} &=& \left\{
\left\{ \begin{array}{c}
           e_1e_2e_3e_4e_5e_6[OK=true]e_7e_8,\\
           e_1e_2e_3e_4e_5e_6[OK=false] \end{array}
 \right\}\right\}\\
 &=& \{\{r_1,r_2\}\}
\end{eqnarray*} where $r_1$ and $r_2$ are given in Example~\ref{ex:behavior}. Recall from Example~\ref{ex:running:semantics},
$\sem{\mathit{Login}} =
\{\{r_1,r_2\},\{r_1',r_2'\}\}$ where $r_1'$ and $r_2'$ are
also given in Example~\ref{ex:behavior}.

Put $D_1=Login2$ and $D_2=Login$. Then it can be easily checked
that the condition 2 in Definition~\ref{df:sd-conform} holds. The
condition 1 in Definition~\ref{df:sd-conform} holds because
$\rho(f_i)=f_i$ and $f_i\not\in\events(Login)$ for $9\leq i\leq
12$. Thus, SD Login2 conforms to SD Login with respect to $\rho$ and
$\mathcal{U}$.
\end{example}

\section{Case example: Mandatory Access Control} \label{sec:case}

This section illustrates via an example how the conformance of an SD
to an access control pattern can be verified.  Access control is an
important aspect in trustworthiness computing to ensure integrity,
confidentiality and availability of shared resources in a
system. Thus, their behaviors must be strictly observed, otherwise
security breaches or denial of services to authorized users may
occur. We use Mandatory Access Control (MAC) \cite{SS94} which governs
access based on security levels.

\begin{figure*}
\begin{center}
\scalebox{0.40}{\includegraphics{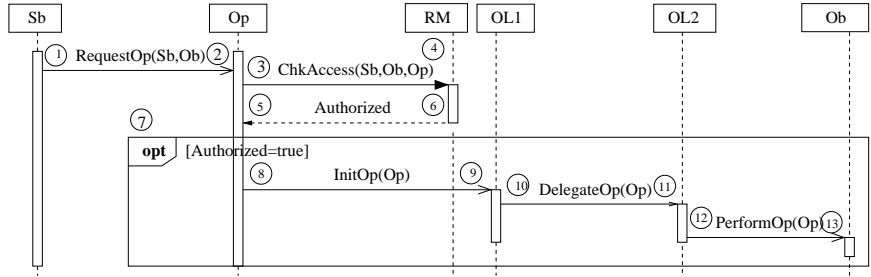}}
\end{center}
\caption{An Instance of MAC Interaction Pattern}
\label{fig:mac-instance}
\end{figure*}

Figure~\ref{fig:mac-instance} shows the interaction behavior of MAC.
The SD describes that subject \textit{Sb} requests operation
\textit{Op} to be performed on object \textit{Ob}. The request is
checked for accessibility by the ChkAccess operation on reference
monitor \textit{RM} which enforces the {\em Simple Security property}
and the {\em restricted-* property} \cite{SS94} for controlling read
and write accesses.  The \textit{opt} fragment specifies that if the
access is authorized, the request is sent to the target object through
two object liaisons \textit{OL1} and \textit{OL2} which delegate the
request.  The SD is represented in the abstract syntax \( D_{Inst}=
block(\{1..7\}, \iota, \{[(i,i+1) \mid 1\leq i\leq 6\}) \) where
\begin{eqnarray*}
\iota(1) & = & (!,RequestOp(Sb,Ob),Sb,Op)\\
\iota(2) &=& (?,RequestOp(Sb,Ob),Sb,Op)\\
\iota(3) &=& (!,ChkAccess(Sb,Ob,Op),Op,RM)\\
\iota(4) &=& (?,ChkAccess(Sb,Ob,Op),Op,RM)\\
\iota(5) &=& (!,Authorized,RM,Op)\\
\iota(6) &=& (?,Authorized,RM,Op)\\
\iota(7) &=&
 opt(Authorized=true,\\
&&   ~~~~~block(\{8..13\},\iota_1,
    \{(i,i+1) \mid 8\leq i\leq 12\})\\
 \iota_1(8) &=& (!,InitOp(Op),Op,OL1)\\
 \iota_1(9) &=& (?,InitOp(Op),Op,OL1)\\
 \iota_1(10) &=& (!,DelegateOp(Op),OL1,OL2)\\
\iota_1(11)  &=& (?,DelegateOp(Op),OL1,OL2)\\
\iota_1(12) &=& (!,PerformOp(Op),OL2,Obj)\\
\iota_1(13) &=& (?,PerformOp(Op),OL2,Obj)\\
\end{eqnarray*}

\add{We have developed a prototype tool for conformance inference in
  Prolog.  Given two SDs $D_1$ and $D_2$, the tool finds every pair
  $\langle \mathcal{U},\rho\rangle$ such that  $D_1\conformsto_{\rho,\mathcal{U}}D_2$. The tool infers that $D_{App}$ in
  Example~\ref{ex:mac2} conforms to $D_{Inst}$ with respect to $\rho$ and
  $\mathcal{U}$ given below. Since $D_{Inst}$ is an instance of the
  MAC pattern, we conclude that $D_{App}$ conforms to the MAC
  pattern.}

\[\mathcal{U}= \left\{
\begin{array}{l}
(!,sort(receiver), sorter, sorter), \\
(?,sort(receiver), sorter, sorter),\\
(!, log(receiver), deliver,transaction), \\
(?, log(receiver), deliver,transaction)
\end{array}\right\}
\]
\[ \rho=\left\{
\begin{array}{l}
 request \mapsto RequestOp, check \mapsto ChkAccess,\\ perform \mapsto
 InitOp, send2 \mapsto DelegateOp,\\ send3 \mapsto PerformOp,\\
 authorized\mapsto Authorized,\\ found\mapsto true, sender \mapsto
 Sb, op \mapsto Op, receiver \mapsto Ob,\\ sl \mapsto RM, sorter
 \mapsto OL1, deliver \mapsto OL2
\end{array}
\right\}
\]

\section{Case Example: JHotDraw} \label{sec:case2}
We have also conducted a case study using JHotDraw 5.2, an open source
framework for building graphical drawing editors. JHotDraw is known to
be pattern-based where sixty instances of ten different design
patterns are found \cite{TCSH06}. We specifically looked into the
three instances of the Observer pattern. We reverse-engineered the
instances using NetBeans 5.5 to generate corresponding UML sequence
diagrams and combined them to make the pattern behavior more
explicit. We checked conformance relationship between the combined
sequence diagram and  the Observer IPS presented in our previous work
\cite{Kim07}.

\begin{figure*}
\begin{center}
\scalebox{0.35}{\includegraphics{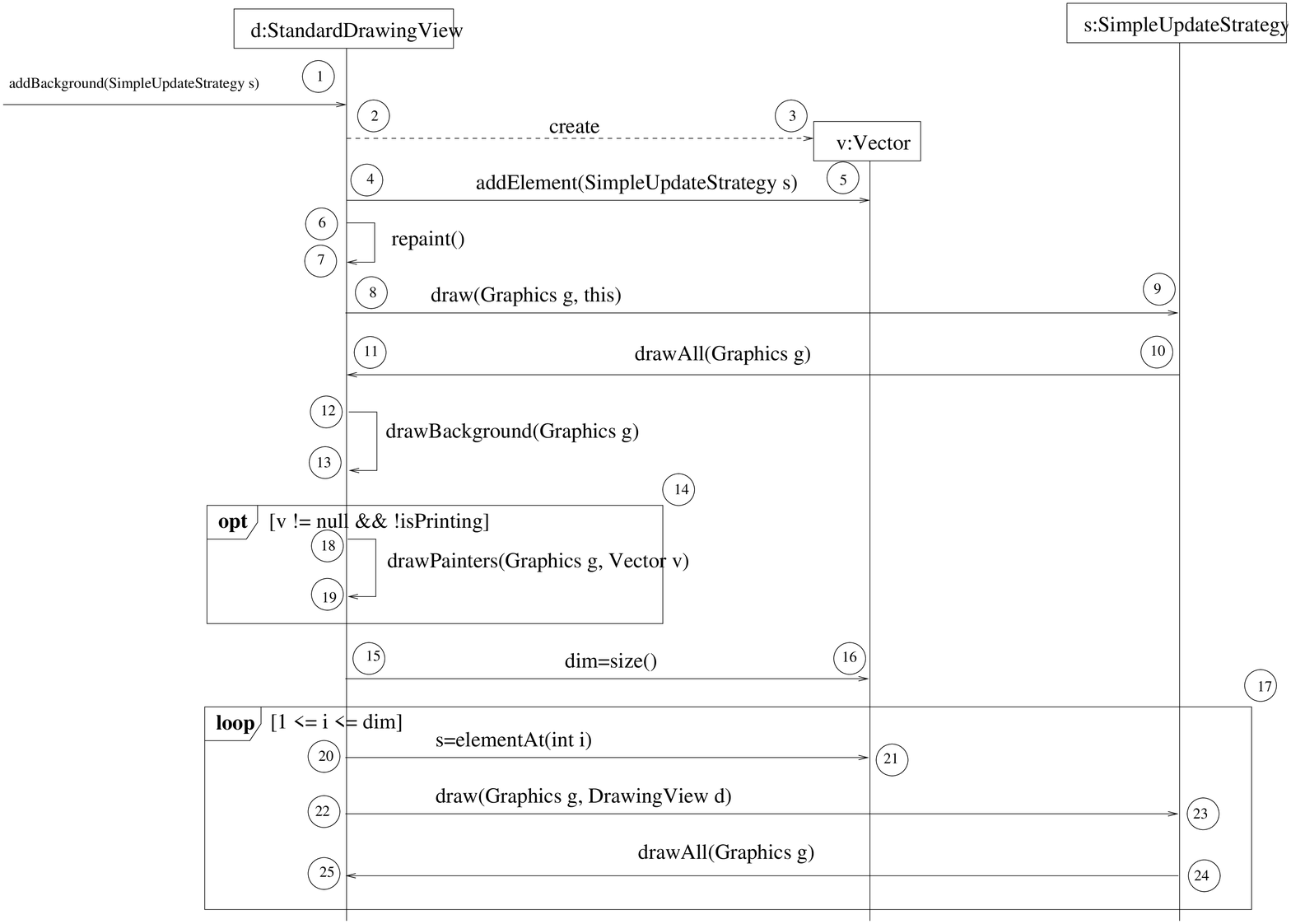}}
\end{center}
\caption{SD for JHotDraw}
\label{fig:jhotdraw}
\end{figure*}

The SD in Fig.~\ref{fig:jhotdraw} describes the part of JHotDraw
behavior that pertains to adding backgrounds to a drawing view through
painters which defines the interface for drawing a layer into the
view. When there is a request for adding a background as an instance
of a painter implementation, the requested background is stored in a
vector and the current view is repainted for each background in the
vector. We have labelled events and combined fragments.  For instance,
the \textit{opt} fragment is labelled $14$. The sending and receiving
events for a message in the SD are labelled with two consecutive
numbers. Let $e_i$ abbreviate the event labelled $i$. For instance,
$e_2$ abbreviates $(!,create(),d,v)$. Then the SD is expressed as \(
D_{jhd} = block( \{ 1..17\}, \{i\mapsto e_i\mid 1\leq i \leq 17
\wedge i\neq 14\wedge i\neq 17\} \cup\{14\mapsto f_{14}, 17\mapsto
f_{17}\}, \strict_0)\) where $\strict_0=\{\langle i,{i+1}\rangle\mid
1\leq i\leq 16\}$. The sub-SD $f_{14}= opt(v!=null \wedge !isPrinting,
block(\{18,19\}, \{i\mapsto e_i \mid 18\leq i\leq 19\}, \{\langle
18,19\rangle\}))$ and the sub-SD $f_{17}=loop( (1\leq i)\wedge (i\leq
dim), block(\{20..25\},\{j\mapsto e_j\mid 20\leq j\leq 25\},\{\langle
e_j,e_{j+1}\rangle \mid 20\leq j <25\}))$.

\begin{figure*}
\begin{center}
\scalebox{0.45}{\includegraphics{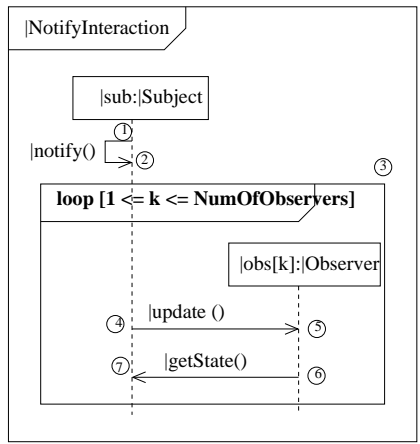}}
\end{center}
\caption{SD for Observer pattern}
\label{fig:visitor}
\end{figure*}

The SD for the Observer pattern is shown in
Fig.~\ref{fig:visitor}. Abbreviate the event labelled with $j$ as
$e_j'$. Then the SD is represented as $D_{obs}= block(\{1,2,3\},$ $
\{1\mapsto e_1', 2\mapsto e_2', 3\mapsto f_3\},\{\langle
1,2\rangle,\langle 2,3\rangle\})$ where $f_3=loop(1\leq k\wedge$ $
k\leq NumOfObservers, block(\{4..7, \{i\mapsto e_i'\mid 4\leq
i\leq 7\}\},\{\langle i,i+1\rangle \mid 4\leq i <7\} ))$.

The prototype tool found the following three pairs of values for
$\langle \rho,\mathcal{U}\rangle$ with respect to which $D_{jhd}$
conforms to $D_{obs}$ where non-null indicates any value which is
not null.
{\setlength\arraycolsep{0.1em}
\begin{eqnarray*}
\mathcal{U}_1 &=& \{e_i \mid i\in \{1..5, 8..13, 15,16,18..21\}\} \\
\rho_1        &=&
    \left\{\begin{array}{c} d\mapsto sub, s\mapsto obs,
            repaint\mapsto notify,\\
             draw\mapsto update, drawAll\mapsto getState,\\
              i\mapsto k, dim\mapsto NumOfObservers \end{array}\right\}\\
\mathcal{U}_2&=& \{e_i\mid i\in \{1..11,15,16,18..21\}\}\\
\rho_2        &=&  \left\{\begin{array}{c} d\mapsto sub,s\mapsto
obs, drawBackground\mapsto notify,\\
                    draw\mapsto update, drawAll\mapsto getState,\\ i\mapsto k, dim\mapsto NumOfObservers  \end{array}\right\}\\
\mathcal{U}_3 &=& \{e_i\mid i\in \{1..13,15,16,20,21\}\} \\
\rho_3        &=& \left\{\begin{array}{c} d\mapsto sub,s\mapsto
obs, drawPainters\mapsto notify,\\
                    draw\mapsto update, drawAll\mapsto getState,i\mapsto k,\\ dim\mapsto NumOfObservers,\\
                    isPrinting\mapsto false, v\mapsto \mbox{non-null} \end{array}\right\}
\end{eqnarray*}
}

These three pairs of values correspond to three ways in which
$D_{jhd}$ conforms to $D_{obs}$ and they differ in how notify message is
realized.  Without semantic information about operations in SDs, the
tool cannot tell which of the three ways is intended by the
designer. Nevertheless, information the tool provides is valuable in
that it presents all possible ways the Observer pattern $D_{obs}$ is
realized in $D_{jhd}$. Note that values of $v$ and $isPrinting$ in
$\rho_3$ satisfy the condition of the \texttt{opt} fragment in
$D_{jhd}$.  We have also studied another variant of the Observer
pattern in which the update message carries the object $sub$ as an
argument. The tool infers that $D_{jhd}$ does not conform to the
variant. This is correct since the draw message in $D_{jhd}$ does not
carry the subject $d$ as an argument and $d$ is the lifeline in
$D_{jhd}$ which corresponds to $sub$ in the variant of the Observer
pattern.  The variant of the Observer pattern is an example of
over-specification.

\section{Conclusion and Future Work} \label{sec:conclusion}
Reasoning about conformance between SDs with respect to their
required behavior is an important issue in software development
process such as aspect-oriented and pattern-based software
development. In this paper, we have presented a trace semantics for
SDs that captures precisely required behavior of SDs and formalized a
notion of conformance based on the semantics. By way of two case examples,
we showed how pattern conformance can be verified.

One future work will be integrating class diagram conformance
presented in~\cite{LuKZK10} and SD conformance relation presented
in this paper.  Another future work is to extend the semantics and
the conformance relation to include interaction operators
\textit{neg} and \textit{assert}.  This requires to take into
account the proscribed behaviors of SDs.  We also plan to use the
semantics proposed in this paper as a basis to investigate the
correctness of the algorithms that translate SDs to other design
models such as statecharts and modal transitions systems.

\appendix

\section{Proofs}
\proofoflemma{lm:one} {By structural induction on $D$. The base cases where $D=\tau$ and $D=e$ are trivial.
Assume that $D=opt(c,D_1)$, by induction hypothesis,
$\exists\mathcal{O}_1\in\sem{D_1}.(\mathcal{O}_1\neq\emptyset)$.
Let $\mathcal{O}= \{ct \mid t \in \mathcal{O}_1\}\cup
\{\neg_{c}(c)\}$. We have that $\mathcal{O}\neq \emptyset$ and
$\mathcal{O}\in \sem{D}$. Other inductive cases are similar.
}

\proofoflemma{lm:two} {Proof can be done by structural induction on $D$  in a similar
way to Lemma~\ref{lm:one} except that there is only one base case
since $\events(e)\neq\emptyset$.
}

\proofofprop{prop} {The proof is done by structural induction on $C$.
In the base case, $C=\mathbbm{x}$. We have $\sem{C[D_1]}=
\sem{D_1}=\sem{D_2}= \sem{C[D_2]}$.

Now assume that $C=opt(c,C')$. By the induction hypothesis, we
have  $\sem{C'[D_1]}= \sem{C'[D_2]}$. Then
\begin{eqnarray*}
\sem{C[D_1]} &=& \sem{opt(c,C'[D_1])}\\
& = & \downarrow\mathit{fold}(\{\{c\}\}\bullet^\sharp
\sem{C'[D_1]} \cup^\sharp \{\{\neg_{c}(c)\}\})\\
& = & \downarrow\mathit{fold}(\{\{c\}\}\bullet^\sharp
\sem{C'[D_2]} \cup^\sharp \{\{\neg_{c}(c)\}\})\\
&=& \sem{opt(c,C'[D_2])}\\
&=& \sem{C[D_2]}
\end{eqnarray*}
Other inductive cases are similar.
}

\proofoflemma{lm:aux1} {Without loss of generality, we assume that  $\alpha_1$ is a sequence of tokens,   for otherwise the result follows
immediately. Then by definition of $\ltimes $, there are an $\alpha'$
such that $\alpha_1\curvearrowright^\ast\alpha'$ and an
$\eta:\dom(\alpha')\mapsto\dom(\alpha_2)$ such that
\begin{itemize}
 \item[(a)] For any $i\in\dom(\alpha')$, $\alpha'(i)\ltimes \alpha_2(\eta(i))$; and
\item[(b)] For any $j\in\dom(\alpha_2)$, if
$j\not\in\mathit{image}(\eta)$ then $\alpha_2(j)\in\constraints$.
\end{itemize}
Since $\alpha_2\curvearrowright \beta_2$, there are
$\omega_1,\omega_2$ and $\omega_3$ such that
$\alpha_2=\omega_1\region{\omega_2}\omega_3$ and
$\beta_2=\omega_1\omega_2\omega_3$. Since $\eta$ is strictly
increasing and $\alpha_2(\|\omega_1\|) =
\region{\omega_2}\not\in\constraints$,
 there is a unique $\ell$ such that $\eta(\ell)=\|\omega_1\|$.
There are also $u_1,u_2$ and $u_3$ such that $\|u_1\|=\ell$,
$\alpha'=u_1\region{u_2}u_3$ and
$u_2\ltimes \omega_2$. Let $\alpha''=u_1u_2u_3$. Then
 $\alpha\curvearrowright^\ast\alpha'\curvearrowright \alpha''$.
Since $u_2$ and $\omega_2$ do not contain critical segment tokens
and $u_2\ltimes \omega_2$, there is an
$\eta':\dom(u_2)\mapsto \dom(\omega_2)$ such that
\begin{itemize}
 \item[(c)] For any $i'\in\dom(u_2)$, $u_2(i')\ltimes \omega_2(\eta'(i'))$; and
\item[(d)] For any $j'\in\dom(\omega_2)$, if
$j'\not\in\mathit{image}(\eta')$ then
$\omega_2(j')\in\constraints$.
\end{itemize}
Since $\alpha_1\curvearrowright^\ast u_1\region{u_2}u_3$, there
are $v_1$ and $v_2$ such that $\alpha_1=v_1\region{u_2}v_3$ and
$v_1 \curvearrowright^\ast u_1$ and $v_3 \curvearrowright^\ast
u_3$. Let $\beta_1=v_1u_2v_3$. Then $\alpha_1\curvearrowright
\beta_1\curvearrowright^\ast u_1u_2u_3=\alpha''$. Now define
$\eta'':\dom(\alpha'')\mapsto \dom(\beta_2)$ as follows.
\[ \eta''(i'') = \left\{\begin{array}{ll}
\eta(i'') & i''<\ell \\
\eta'(i''-\ell) & \ell\leq i'' < \ell+\|\omega_2\|\\
\eta(i''-\|\omega_2\|+1) & i''\geq \ell+\|\omega_2\|
\end{array}\right.
\]
The following follows from (a)-(d).
\begin{itemize}
 \item For any $i''\in\dom(\alpha'')$, $\alpha''(i'')\ltimes \beta_2(\eta''(i''))$; and
\item For any $j''\in\dom(\beta_2)$, if
$j''\not\in\mathit{image}(\eta'')$ then
$\beta_2(j'')\in\constraints$.
\end{itemize}
  So,
$\beta_1\ltimes \beta_2$.
}

\proofoflemma{lm:aux2} {Let $\gamma$ be an arbitrary trace.
We prove $\gamma\ltimes \gamma$ by structural induction on
$\gamma$. In the base case where $\gamma=e$, $e\ltimes e$ by
definition. In the base case where $\gamma=c$, $c\ltimes c$ follows
from reflexivity of $\models_{c}$. In the case where
$\gamma=\region{\gamma'}$, we have $\gamma'\ltimes \gamma'$ by the
induction hypothesis, which implies $\gamma \ltimes \gamma$.  Assume
that $\gamma=t_1\cdots t_n$. By the induction hypothesis, we have that
$t_i \ltimes t_i$ for $0\leq i < n$.  Then $\gamma \ltimes \gamma$ by
putting $\alpha=\beta=\gamma$ and $\eta(i)=i$ for any $0\leq i <n$ in
the definition of $ \ltimes $.

 Assume that $\alpha \ltimes \beta$ and $\beta \ltimes \gamma$. We
 prove $\alpha\ltimes \gamma$ by structural induction on $\alpha$.
 Case (a): $\alpha=e$. Then $\beta$ must be of the form $\omega_1 e
 \omega_2$ with $\omega_1,\omega_2\in\constraints^\ast$. Since $\beta
 \ltimes \gamma$, there is a strictly increasing function
 $\eta':\dom(\beta)\mapsto\dom(\gamma)$ such that
\begin{itemize}
 \item[(1')] For any $i\in\dom(\beta)$, $\beta(i)\ltimes
 \gamma(\eta'(i))$;

\item[(2')] For any $j\in\dom(\gamma)$, if
$j\not\in\mathit{image}(\eta')$ then $\gamma(j)\in\constraints$;
\end{itemize}
Let $\ell$ be the unique position at which $e$ occurs in $\beta$.
Then (2') implies that $\gamma(j)\in\constraints$ for all $j\neq
\eta'(\ell)$; (1') implies that $\gamma(\eta'(\ell))=e$.  Thus,
$\alpha\ltimes \gamma$.

Case (b): $\alpha=c$ for some $c\in\constraints$. Similar to Case
(a).

Case (c): $\alpha=\region{\alpha'}$. There are $\beta'$ and
$\gamma'$ such that $\beta=\region{\beta'}$,
$\gamma=\region{\gamma'}$, $\alpha'\ltimes  \beta'$
and $\beta'\ltimes \gamma'$. By the induction
hypothesis, we have $\alpha' \ltimes \gamma'$ which
implies $\alpha \ltimes \gamma$.

Case (d): Since $\beta\ltimes \gamma$, there are
$\beta'$ such that $\beta\curvearrowright^\ast\beta'$ and
$\eta_2:\dom(\beta')\mapsto\dom(\gamma)$ such that
\begin{itemize}
 \item[(i)] For any $i\in\dom(\beta')$, $\beta'(i)\ltimes
 \gamma(\eta_2(i))$;

\item[(ii)] For any $j\in\dom(\gamma)$, if
$j\not\in\mathit{image}(\eta_2)$ then $\gamma(j)\in\constraints$;
\end{itemize}
Since $\alpha\ltimes \beta$, there is an $\alpha'$
such that $\alpha\curvearrowright^\ast\alpha'$ and
$\alpha'\ltimes \beta'$ by Lemma~\ref{lm:aux1}. Thus,
there are an $\alpha''$ such that $\alpha' \curvearrowright^\ast
\alpha''$ and an $\eta_1:\dom(\alpha'')\mapsto \dom(\beta')$ such
that
\begin{itemize}
 \item[(iii)] For any $i\in\dom(\alpha'')$, $\alpha''(i)\ltimes
 \beta'(\eta_1(i))$;

\item[(iv)] For any $j\in\dom(\beta')$, if
$j\not\in\mathit{image}(\eta_1)$ then $\beta'(j)\in\constraints$;
\end{itemize}
Now define $\eta:\dom(\alpha'')\mapsto\dom(\gamma)$ by $\eta =
\eta_2\circ\eta_1$. Then (i)-(iv) imply that
\begin{itemize}
 \item For any $i\in\dom(\alpha'')$, $\alpha''(i)\ltimes
 \gamma(\eta(i))$;

\item For any $j\in\dom(\gamma)$, if
$j\not\in\mathit{image}(\eta)$ then $\gamma(j)\in\constraints$;
\end{itemize}
This, together with $\alpha\curvearrowright^\ast\alpha''$, implies
that $\alpha\ltimes \gamma$.
}

\proofoflemma{remark1} {Let  $hide_{\mathcal{U}}(t)$ be the result of replacing  each occurrence of $e$ in $t$ with $\epsilon$ for each
$e\in\mathcal{U}$.  $hide_{\mathcal{U}}$ is extended to obligations
  and meanings as $\mathit{tag}$ is.  Then
  $\sem{hide_{\mathcal{U}}(D_i)} = hide_{\mathcal{U}}(\sem{D_i})$
  for $i=1,2$. By a simple structural induction on $t_1$, we have
$t_1\ltimes t_2$ implies $hide_{\mathcal{U}}(t_1)\ltimes hide_{\mathcal{U}}(t_2)$. The result follows.
}

\proofoflemma{remark2}{
Observe $\sem{\rho(D_i)} = \rho(\sem{D_i})$ for $i=1,2$. Since $\rho$
is a function on $\names$, we have $t_1\ltimes t_2$ implies
$\rho(t_1)\ltimes \rho(t_2)$. So, the result follows.
}

\proofoftheorem{th:1}{
\begin{enumerate}
\item To prove $D\conformsto  D$. Let $D_1=D_2=D$. Put
  $\mathcal{U}=\emptyset$ and $\rho(x)=x$ for all $x\in\names$. Then
  condition (1) in Definition~\ref{df:sd-conform} holds since
  $\mathcal{U}=\emptyset$. Consider condition (2) in
  Definition~\ref{df:sd-conform}. We have $D_1'=D_2'=D$. Then condition (2)
  in Definition~\ref{df:sd-conform} holds because of reflexivity of $\ltimes
  $.
\item Now assume $D_1 \conformsto  D_2$ and $D_2
  \conformsto  D_3$.  Then there are renaming substitutions
  $\rho_{1},\rho_{2}$ and sets of events
  $\mathcal{U}_1\subseteq\events(D_1),\mathcal{U}_2\subseteq\events(D_2)$
  such that
\begin{itemize}
\item [(a)] $\rho_2(\mathcal{U}_2)\cap \events(D_3)=\emptyset$,

 \item [(b)] \( \forall \mathcal{O}_2\in\sem{D_2'}.\exists
   \mathcal{O}_3\in\sem{D_3}.\forall t_3\in \mathcal{O}_3.\exists
   t_2\in \mathcal{O}_2. (t_2\ltimes t_3) \) where $D_2'=
   \rho_2(hide_{\mathcal{U}_2}(D_2))$.

\item [(c)] $\rho_1(\mathcal{U}_1)\cap \events(D_2)
=\emptyset$, and

\item [(d)] \( \forall \mathcal{O}_1\in\sem{D_1'}.\exists
\mathcal{O}_2\in\sem{D_2}.\forall t_2\in \mathcal{O}_2.\exists
t_1\in \mathcal{O}_1. (t_1\ltimes  t_2) \) where
$D_1'= \rho_1(hide_{\mathcal{U}_1}(D_1))$.
\end{itemize}
Let $\rho=\rho_{2}\circ\rho_{1}$, and $\mathcal{U}=\mathcal{U}_1 \cup \mathcal{U}_1 '$ where $\mathcal{U}_1'=
\{e\mid e\in\events(D_1) \wedge \rho_1(e)\in \mathcal{U}_2\}$. Then
$\mathcal{U}\subseteq\events(D_1)$. Let $e_3$ be an arbitrary event in
$\events(D_3)$ and $e_1$ be an arbitrary event in $\events(D_1)$ such
that $e_1=\rho(e_3)$. We now prove that $e_1\not\in\mathcal{U}$ by way
of contradiction. Assume that $e_1\in\mathcal{U}$. Then there is no
event $e_2\in\events(D_2)$ such that $\rho_1(e_1)=e_2$ according to
(c). Thus, the condition (1) in Definition~\ref{df:sd-conform} holds.

Now consider the condition (2) in Definition~\ref{df:sd-conform}. Note that
$\rho_1(\mathcal{U}_1') = \mathcal{U}_2$.
\begin{eqnarray*}
  D_1'' & = & \rho\circ hide_\mathcal{U} (D_1) \\
 &=& \rho_2\circ \rho_1 \circ hide_{\mathcal{U}_1\cup \mathcal{U}_1'}(D_1)\\
 &=& \rho_2\circ \rho_1 \circ hide_{\mathcal{U}_1'}\circ hide_{\mathcal{U}_1} (D_1)\\
 &=& \rho_2  \circ hide_{\mathcal{U}_2}\circ \rho_1\circ hide_{\mathcal{U}_1} (D_1)\\
&=& \rho_2  \circ hide_{\mathcal{U}_2}(D_1').
\end{eqnarray*}
The condition 2 in Definition~\ref{df:sd-conform} then follows from
Remarks~\ref{remark1} and~\ref{remark2} and transitivity of $\ltimes$.
\end{enumerate}
}

\begin{thebibliography}{10}

\bibitem{AlurEY03}
R.~Alur, K.~Etessami, and M.~Yannakakis.
\newblock Inference of message sequence charts.
\newblock {\em IEEE Trans. Software Eng.}, 29(7):623--633, 2003.

\bibitem{AlurY99}
R.~Alur and M.~Yannakakis.
\newblock Model checking of message sequence charts.
\newblock In J.~C.~M. Baeten and S.~Mauw, editors, {\em Proceedings of 10th
  International Conference on Concurrency Theory}, volume 1664 of {\em Lecture
  Notes in Computer Science}, pages 114--129. Springer, 1999.

\bibitem{Aredo02}
D.~B. Aredo.
\newblock A framework for semantics of {UML} sequence diagrams in {PVS}.
\newblock {\em J. UCS}, 8(7):674--697, 2002.

\bibitem{BaaderT98}
F.~Baader and T.~Nipkow.
\newblock {\em Term rewriting and all that}.
\newblock Cambridge University Press, 1998.

\bibitem{BergstraK85}
J.A. Bergstra and J.W. Klop.
\newblock Algebra of communicating processes with abstraction.
\newblock {\em Theor. Comput. Sci.}, 37:77--121, 1985.

\bibitem{Bontemps:2005:LSC}
Y.~Bontemps, P.~Heymans, and P.-Y. Schobbens.
\newblock From live sequence charts to state machines and back: A guided tour.
\newblock {\em IEEE Trans. Softw. Eng.}, 31:999--1014, 2005.

\bibitem{Broy05}
M.~Broy.
\newblock A semantic and methodological essence of message sequence charts.
\newblock {\em Sci. Comput. Program.}, 54(2-3):213--256, 2005.

\bibitem{CardosoS01}
J.~Cardoso and C.~Sibertin-Blanc.
\newblock {Ordering actions in sequence diagrams of UML}.
\newblock In {\em Proceedings of 23rd International Conference on Information
  Technology Interfaces}, pages 3--14, 2001.

\bibitem{CavarraK04}
A.~Cavarra and J.~K{\"u}ster-Filipe.
\newblock Formalizing liveness-enriched sequence diagrams using asms.
\newblock In W.~Zimmermann and B.~Thalheim, editors, {\em Proceedings of 11th
  International Workshop on Abstract State Machines}, volume 3052 of {\em
  Lecture Notes in Computer Science}, pages 62--77. Springer, 2004.

\bibitem{CengarleGW06}
M.~V. Cengarle, P.~Graubmann, and S.~Wagner.
\newblock Semantics of {UML} 2.0 interactions with variabilities.
\newblock {\em Electr. Notes Theor. Comput. Sci.}, 160:141--155, 2006.

\bibitem{CengarleK04}
M.~V. Cengarle and A.~Knapp.
\newblock {UML} 2.0 interactions: Semantics and refinement.
\newblock In {\em Proceedings of the 3rd Int. Wsh. Critical Systems Development
  with UML}, pages 85--99, 2004.

\bibitem{ChenKS05}
C.A. Chen, S.~Kalvala, and J.~Sinclair.
\newblock A process-based semantics for message sequence charts with data.
\newblock In {\em Australian Software Engineering Conference}, pages 130--139,
  2005.

\bibitem{ChoKCB01}
S.~M. Cho, H.~H. Kim, S.~D. Cha, and D.~H. Bae.
\newblock Specification and validation of dynamic systems using temporal logic.
\newblock {\em IEE Proceedings - Software}, 148(4):135--140, 2001.

\bibitem{ChoKCB02}
S.~M. Cho, H.~H. Kim, S.~D. Cha, and D.~H. Bae.
\newblock A semantics of sequence diagrams.
\newblock {\em Inf. Process. Lett.}, 84(3):125--130, 2002.

\bibitem{CleavelandH93}
R.~Cleaveland and M.~Hennessy.
\newblock Testing equivalence as a bisimulation equivalence.
\newblock {\em Formal Asp. Comput.}, 5(1):1--20, 1993.

\bibitem{DammH01}
W.~Damm and D.~Harel.
\newblock {LSC}s: Breathing life into message sequence charts.
\newblock {\em Formal Methods in System Design}, 19(1):45--80, 2001.

\bibitem{EichnerFMSS05}
C.~Eichner, H.~Fleischhack, R.~Meyer, U.~Schrimpf, and C.~Stehno.
\newblock Compositional semantics for {UML} 2.0 sequence diagrams using {Petri
  Nets}.
\newblock In A.~Prinz, R.~Reed, and J.~Reed, editors, {\em Proceedings of the
  12th International {SDL} Forum}, volume 3530 of {\em Lecture Notes in
  Computer Science}, pages 133--148. Springer, 2005.

\bibitem{Engels98}
A.~Engels.
\newblock Message refinement: Describing multi-level protocols in {MSC}.
\newblock In {\em Proceedings of the 1st Workshop of the SDL Forum Society on
  SDL and MSC, number 104 in Informatik-Berichte}, pages 67--74, 1998.

\bibitem{EshuisF02}
R.~Eshuis and M.~M. Fokkinga.
\newblock Comparing refinements for failure and bisimulation semantics.
\newblock {\em Fundam. Inform.}, 52(4):297--321, 2002.

\bibitem{Fernandes07}
J.~M. Fernandes, S.~Tjell, J.~B. Jorgensen, and O.~Ribeiro.
\newblock Designing tool support for translating use cases and {UML} 2.0
  sequence diagrams into a coloured {Petri Net}.
\newblock In {\em Proceedings of 6th International Workshop on Scenarios and
  State Machines}. IEEE Computer Society, 2007.

\bibitem{FischbeinBU09}
D.~Fischbein, V.~A. Braberman, and S.~Uchitel.
\newblock A sound observational semantics for modal transition systems.
\newblock In M.~Leucker and C.~Morgan, editors, {\em Proceedings of 6th
  International Colloquium Theoretical Aspects of Computing}, volume 5684 of
  {\em Lecture Notes in Computer Science}, pages 215--230. Springer, 2009.

\bibitem{FischbeinU08}
D.~Fischbein and S.~Uchitel.
\newblock On correct and complete strong merging of partial behaviour models.
\newblock In M.~J. Harrold and G.~C. Murphy, editors, {\em Proceedings of 16th
  ACM SIGSOFT International Symposium on Foundations of Software Engineering},
  pages 297--307. ACM, 2008.

\bibitem{FischbeinUB06}
D.~Fischbein, S.~Uchitel, and V.~Braberman.
\newblock A foundation for behavioural conformance in software product line
  architectures.
\newblock In {\em Proceedings of the ISSTA 2006 workshop on Role of software
  architecture for testing and analysis}, pages 39--48. The ACM Press, 2006.

\bibitem{FranceRGG04}
R.~France, I.~Ray, G.~Georg, and S.~Ghosh.
\newblock An aspect-oriented approach to design modeling.
\newblock {\em IEE Proceedings - Software}, 151(4):173--–185, 2004.

\bibitem{GHJV95}
E.~Gamma, R.~Helm, R.~Johnson, and J.~Vlissides.
\newblock {\em Design Patterns: Elements of Reusable Object-Oriented Software}.
\newblock Addison Wesley, 1995.

\bibitem{GiannakopoulouM03}
D.~Giannakopoulou and J.~Magee.
\newblock Fluent model checking for event-based systems.
\newblock In {\em Proceedings of 9th European Software Engineering Conference},
  pages 257--266. ACM, 2003.

\bibitem{GrosuS05}
R.~Grosu and SA~Smolka.
\newblock {Safety-liveness semantics for UML 2.0 sequence diagrams}.
\newblock In {\em Proceedings of 5th International Conference on Application of
  Concurrency to System Design}, pages 6--14. IEEE Computer Society, 2005.

\bibitem{GSJ00}
A.L. Guennec, G.~Sunye, and J.~Jezequel.
\newblock {Precise Modeling of Design Patterns}.
\newblock In {\em Proceedings of UML'00}, pages 482--496, 2000.

\bibitem{Hammal06}
Y.~Hammal.
\newblock Branching time semantics for {UML} 2.0 sequence diagrams.
\newblock In {\em Proceedings of 26th IFIP WG 6.1 International Conference on
  Formal Techniques for Networked and Distributed Systems}, volume 4229 of {\em
  Lecture Notes in Computer Science}, pages 259--274. Springer, 2006.

\bibitem{HarelKM07}
D.~Harel, A.~Kleinbort, and S.~Maoz.
\newblock {S2A}: A compiler for multi-modal uml sequence diagrams.
\newblock In M.~B. Dwyer and A.~Lopes, editors, {\em Proceedings of 10th
  International Conference on Fundamental Approaches to Software Engineering},
  volume 4422 of {\em Lecture Notes in Computer Science}, pages 121--124.
  Springer, 2007.

\bibitem{HarelM08}
D.~Harel and S.~Maoz.
\newblock Assert and negate revisited: Modal semantics for uml sequence
  diagrams.
\newblock {\em Software and System Modeling}, 7(2):237--252, 2008.

\bibitem{HarelMS08}
D.~Harel, S~Maoz, and I~Segall.
\newblock Some results on the expressive power and complexity of {LSC}s.
\newblock In A.~Avron, N.~Dershowitz, and A.~Rabinovich, editors, {\em Pillars
  of Computer Science, Essays Dedicated to Boris (Boaz) Trakhtenbrot on the
  Occasion of His 85th Birthday}, volume 4800 of {\em Lecture Notes in Computer
  Science}, pages 351--366. Springer, 2008.

\bibitem{Harel:2003}
David Harel and Rami Marelly.
\newblock {\em Come, Let's Play: Scenario-Based Programming Using LSC's and the
  Play-Engine}.
\newblock Springer-Verlag New York, Inc., Secaucus, NJ, USA, 2003.

\bibitem{HaugenHRS05}
{\O}.~Haugen, K.~E. Husa, R.~K. Runde, and K.~St{\o}len.
\newblock {STAIRS} towards formal design with sequence diagrams.
\newblock {\em Software and System Modeling}, 4(4):355--367, 2005.

\bibitem{Hennessy85}
M.~Hennessy.
\newblock Acceptance trees.
\newblock {\em J. ACM}, 32(4):896--928, 1985.

\bibitem{Hennessy88}
M.~Hennessy.
\newblock {\em Algebraic Theory of Processess}.
\newblock The MIT Press, 1988.

\bibitem{Hoa85}
C.~A.~R. Hoare.
\newblock {\em Communicating Sequential Processes}.
\newblock Prentice-Hall, 1985.

\bibitem{Khendek:2001:SDM}
F.~Khendek, S.~Bourduas, and D.~Vincent.
\newblock Stepwise design with message sequence charts.
\newblock In {\em Proceedings of the IFIP TC6/WG6.1 - 21st International
  Conference on Formal Techniques for Networked and Distributed Systems}, pages
  19--34. Kluwer, B.V., 2001.

\bibitem{Kim07}
D.~Kim.
\newblock {The Role-Based Metamodeling Language for Specifying Design
  Patterns}.
\newblock In Toufik Taibi, editor, {\em Design Pattern Formalization
  Techniques}, pages 183--205. Idea Group Inc., 2007.

\bibitem{KnappW06}
A.~Knapp and J.~Wuttke.
\newblock Model checking of {UML} 2.0 interactions.
\newblock In T.~K{\"u}hne, editor, {\em Models in Software Engineering,
  Workshops and Symposia at MoDELS 2006, Reports and Revised Selected Papers},
  volume 4364 of {\em Lecture Notes in Computer Science}, pages 42--51.
  Springer, 2007.

\bibitem{KohlmeyerG09}
J.~Kohlmeyer and W.~Guttmann.
\newblock Unifying the semantics of uml 2 state, activity and interaction
  diagrams.
\newblock In A.~Pnueli, I.~Virbitskaite, and A.~Voronkov, editors, {\em
  Perspectives of Systems Informatics, 7th International Andrei Ershov Memorial
  Conference, PSI 2009. Revised Papers}, volume 5947 of {\em Lecture Notes in
  Computer Science}, pages 206--217, 2010.

\bibitem{KrkaBEM09}
I.~Krka, Y.~Brun, G.~Edwards, and N.~Medvidovic.
\newblock Synthesizing partial component-level behavior models from system
  specifications.
\newblock In H.~van Vliet and V~Issarny, editors, {\em Proceedings of the 7th
  joint meeting of the European Software Engineering Conference and the ACM
  SIGSOFT International Symposium on Foundations of Software Engineering},
  pages 305--314. ACM, 2009.

\bibitem{LarsenT88}
K.~G. Larsen and B.~Thomsen.
\newblock A modal process logic.
\newblock In {\em Proceedings of 3rd Annual Symposium on Logic in Computer
  Science, Edinburgh, Scotland, UK}, pages 203--210. IEEE Computer Society,
  1988.

\bibitem{LiLJ04}
X.~Li, Z.~Liu, and J.~He.
\newblock A formal semantics of {UML} sequence diagram.
\newblock In {\em Australian Software Engineering Conference}, pages 168--177.
  IEEE Computer Society, 2004.

\bibitem{LuK:ICECCS11}
L.~Lu and D.K. Kim.
\newblock Required behavior of sequence diagrams: Semantics and refinement.
\newblock In {\em Proceedings of the 16th IEEE International Conference on
  Engineering of Complex Computer Systems}. IEEE Computer Society Press, 2011.

\bibitem{LuKZK10}
L.~Lu, D.K. Kim, Y.~Zhu, and S.~Kim.
\newblock Verification of structural pattern conformance using logic
  programming.
\newblock {\em Journal of Universal Computer Science}, 16(17):2455--2474, 2010.

\bibitem{LundS06}
M.~S. Lund and K.~St{\o}len.
\newblock A fully general operational semantics for {UML} 2.0 sequence diagrams
  with potential and mandatory choice.
\newblock In J.~M., T.~Nipkow, and E.~Sekerinski, editors, {\em FM 2006: Formal
  Methods, 14th International Symposium on Formal Methods, Proceedings}, volume
  4085 of {\em Lecture Notes in Computer Science}, pages 380--395. Springer,
  2006.

\bibitem{MHG02}
D.~Mapelsden, J.~Hosking, and J.~Grundy.
\newblock {Design Pattern Modelling and Instantiation using DPML}.
\newblock In {\em Proceedings of the 40th International Conference on
  Technology of Object-Oriented Languages and Systems (TOOLS)}, pages 3--11.
  ACS, 2002.

\bibitem{MauwR94}
S.~Mauw and M.~A. Reniers.
\newblock An algebraic semantics of basic message sequence charts.
\newblock {\em Comput. J.}, 37(4):269--278, 1994.

\bibitem{MauwR96}
S.~Mauw and M.~A. Reniers.
\newblock Refinement in interworkings.
\newblock In U.~Montanari and V.~Sassone, editors, {\em Proceedings of the 7th
  International Conference on Concurrency Theory}, volume 1119 of {\em Lecture
  Notes in Computer Science}, pages 671--686. Springer, 1996.

\bibitem{MicskeiW2010}
Z.~Micskei and H.~Waeselynck.
\newblock The many meanings of uml 2 sequence diagrams: a survey.
\newblock {\em Software and Systems Modeling}, pages 1--26, 2010.
\newblock 10.1007/s10270-010-0157-9.

\bibitem{Milner}
R.~Milner.
\newblock {\em Communication and concurrency}.
\newblock Prentice-Hall, Inc., Upper Saddle River, NJ, USA, 1989.

\bibitem{MuschollP00}
A.~Muscholl and D.~Peled.
\newblock Analyzing message sequence charts.
\newblock In E.~Sherratt, editor, {\em SAM 2000, 2nd Workshop on SDL and MSC},
  pages 3--17. VERIMAG, IRISA, SDL Forum, 2000.

\bibitem{MuschollPS98}
A.~Muscholl, D.~Peled, and Z.~Su.
\newblock Deciding properties for message sequence charts.
\newblock In M.~Nivat, editor, {\em Proceedings of 1st International Conference
  on Foundations of Software Science and Computation Structure}, volume 1378 of
  {\em Lecture Notes in Computer Science}, pages 226--242. Springer, 1998.

\bibitem{NicolaH84}
R.~De Nicola and M.~Hennessy.
\newblock Testing equivalences for processes.
\newblock {\em Theor. Comput. Sci.}, 34:83--133, 1984.

\bibitem{Park81}
D.M.R. Park.
\newblock Concurrency and automata on infinite sequences.
\newblock In P.~Deussen, editor, {\em Proceedings of 5th GI-Conference on
  Theoretical Computer Science}, volume 104 of {\em Lecture Notes in Computer
  Science}, pages 167--183. Springer, 1981.

\bibitem{POMSET}
V.~Pratt.
\newblock The pomset model of parallel processes: Unifying the temporal and the
  spatial.
\newblock Technical Report STAN-CS-85=1049, Department of Computer Science,
  Stanford University, Stanford, CA., 1985.

\bibitem{SS94}
R.~Sandhu and P.~Samarati.
\newblock Access control: Principles and practice.
\newblock {\em IEEE Communications}, 32(9):40--48, 1994.

\bibitem{SenguptaC06}
B.~Sengupta and R.~Cleaveland.
\newblock Triggered message sequence charts.
\newblock {\em IEEE Trans. Software Eng.}, 32(8):587--607, 2006.

\bibitem{SibayUB08}
G.~Sibay, S.~Uchitel, and V.~A. Braberman.
\newblock Existential live sequence charts revisited.
\newblock In W.~Sch{\"a}fer, M.~B. Dwyer, and V.~Gruhn, editors, {\em
  Proceedings of 30th International Conference on Software Engineering}, pages
  41--50. ACM, 2008.

\bibitem{Storrle:assert}
H.~Storrle.
\newblock {Assert, Negate and Refinement in UML-2 Interactions}.
\newblock In {\em Proceedings of 2nd Int. Wsh. Critical Systems Development
  with UML}, pages 79--94, 2003.

\bibitem{Storrle03}
H.~St\"{o}rrle.
\newblock {Semantics of Interactions in UML 2.0}.
\newblock In {\em Proceedings of 2003 IEEE Symposium on Human Centric Computing
  Languages and Environments}, pages 129--136, 2003.

\bibitem{UML05}
{The Object Management Group}.
\newblock {Unified Modeling Language: Superstructure}.
\newblock Version 2.0, OMG Document: formal/2005-07-04, 2005.

\bibitem{TCSH06}
N.~Tsantalis, A.~Chatzigeorgiou, G.~Stephanides, and S.T. Halkidis.
\newblock Design pattern detection using similarity scoring.
\newblock {\em IEEE Transactions on Software Engineering}, 32(11):896--909,
  2006.

\bibitem{UchitelBC09}
S.~Uchitel, G.~Brunet, and M.~Chechik.
\newblock Synthesis of partial behavior models from properties and scenarios.
\newblock {\em IEEE Trans. Software Eng.}, 35(3):384--406, 2009.

\bibitem{UchitelKM03}
S.~Uchitel, J.~Kramer, and J.~Magee.
\newblock Synthesis of behavioral models from scenarios.
\newblock {\em IEEE Trans. Software Eng.}, 29(2):99--115, 2003.

\bibitem{UchitelKM04}
S.~Uchitel, J.~Kramer, and J.~Magee.
\newblock Incremental elaboration of scenario-based specifications and behavior
  models using implied scenarios.
\newblock {\em ACM Trans. Softw. Eng. Methodol.}, 13(1):37--85, 2004.

\bibitem{MSC}
International~Telecommunication Union.
\newblock {\em Message Sequence Chart (MSC)}.
\newblock ITU-T Recommendation Z.120, 1999.

\bibitem{WhittleS00}
J.~Whittle and J.~Schumann.
\newblock Generating statechart designs from scenarios.
\newblock In {\em Proceedings of the 2000 International Conference on Software
  Engineering}, pages 314--323, 2000.

\bibitem{Xiang:2009}
Z.~Xiang and Z.~Shao.
\newblock Asm semantic modeling and checking for sequence diagram.
\newblock In {\em Proceedings of 5th International Conference on Natural
  Computation - Volume 05}, pages 527--530. IEEE Computer Society, 2009.

\bibitem{ZiadiHJ04}
T.~Ziadi, L.~H{\'e}lou{\"e}t, and J.-M. J{\'e}z{\'e}quel.
\newblock Revisiting statechart synthesis with an algebraic approach.
\newblock In {\em Proceedings of 26th International Conference on Software
  Engineering}, pages 242--251. IEEE Computer Society, 2004.

\end{thebibliography}
\end{document}